\documentclass[superscriptaddress,aps,amsmath,amssymb,showpacs,showkeys]{revtex4-2}
\usepackage{natbib} 
\usepackage[dvips]{graphicx}
\usepackage{times}
\usepackage{braket}
\usepackage{xcolor}
\usepackage{orcidlink}
\usepackage{hyperref}
\usepackage{booktabs}
\usepackage{subfigure}
\usepackage{siunitx}
\hypersetup{
	colorlinks=true,
	urlcolor=magenta,
	linkcolor=red,
	citecolor=blue
}

\begin{document}
\title{Galaxy Evolution in the Local Universe: Group Richness Effects on Mergers and Non-Mergers}
\author{Pius Privatus\orcidlink{0000-0002-6981-717X}}
\email[Email: ]{privatuspius08@gmail.com}
\affiliation{Department of Physics, Dibrugarh University, Dibrugarh 786004, Assam, India}
\affiliation{Department of Natural Sciences, Mbeya University of Science and Technology, Iyunga 53119, Mbeya, Tanzania}
\author{Umananda Dev Goswami\orcidlink{0000-0003-0012-7549}}
\email[Email: ]{umananda@dibru.ac.in}
\affiliation{Department of Physics, Dibrugarh University, Dibrugarh 786004, Assam, India}
	
\begin{abstract}
In this study, we use the luminous volume-limited samples obtained from the 
twelfth release of Sloan Digital Sky Survey data and mergers from Galaxy Zoo 
Project to investigate the influence of group richness in shaping galaxy 
properties' distributions and their relationships in the local Universe by 
comparison of mergers and non-mergers. The galaxies were restricted into 
mass-limited subsamples of low-mass, intermediate-mass and high-mass, assigned 
into groups from poor to rich group systems, where the distributions of 
star formation rate (SFR), specific SFR (SSFR), spectral index D$_n$ (4000) and 
$u-r$ colour properties between mergers and non-mergers 
for all subsamples and their relations with stellar mass of galaxies are 
compared. The study revealed a significant difference in the distributions 
between mergers' and non-mergers' properties for low-mass galaxies, while for 
high-mass galaxies the difference is very weak. For the low-mass sample, 
mergers possess higher SFR, SSFR than non-mergers when the group richness is 
kept constant, while for high-mass poor group galaxies have higher SFR, SSFR 
than rich group galaxies when merging status is kept constant. Mergers resemble 
young stellar populations and are bluer than non-mergers for low-mass, while 
for high-mass, mergers and non-mergers have comparable SFR, SSFR, D$_n$ (4000), 
and $u-r$ colour. The study concludes that group richness and stellar mass 
influence the mergers' and non-mergers' properties' distributions, and their 
relationships.
\end{abstract}
	
\maketitle    
	
\section{Introduction} \label{secI}
Galactic interactions and mergings are directly related to questions 
regarding how galaxies form and evolve, how dark matter is distributed, 
and how morphology transforms \cite{ball2008galaxy,van2008dependence,
park2007environmental,springel2000modelling,cox2006feedback,kapferer2005star,
meza2003simulations,lambas2003galaxy,li2008interaction}. For example, a galaxy 
merger may result in the morphological transformation from disk to elliptical, 
which is also expected to influence the star formation rate (SFR) 
\cite{pearson2022north}. Furthermore, galaxy mergers result in the movement 
of materials to the centre of the galaxy, which could result in an increase 
in active galactic nucleus (AGN) activities, despite the fact that the 
connection between AGN and mergers is still a matter of debate 
\cite{bickley2023agns,silva2021galaxy,gao2020mergers}. These effects make 
the interactions and mergings significant factors in the formation and 
evolution of galaxies.

The environment is anticipated to be another important factor shaping the 
formation and evolution of galaxies \cite{bitsakis2011mid,boselli2008origin,kewley2006metallicity,ellison2008galaxy,martin2005galaxy}. 
Ref.~\cite{bitsakis2011mid}, performed a comprehensive study on how the 
compact galaxy groups influences galaxies' evolution, SFRs, and morphological 
transformations using multi-wavelength analysis in the ultraviolet and 
infrared regimes for a sample of 32 Hickson compact groups (HCGs) having 135 
galaxies from Galaxy Evolution Explorer (GALEX) and Spitzer observations 
detailed in Ref.~\cite{martin2005galaxy}. They revealed a significant 
difference in the properties of galaxies based on their group dynamics and 
interactions. Further observed that the evolution of properties with group 
environments is not an instantaneous process supporting the idea that 
environmental mechanisms operates over cosmological timescales, i.e., the 
galaxy transformation is an extended process. How the cluster of galaxies is 
influenced was similar to the binaries in the field environment, though over 
time, the effects of small-scale mergers determine the star formation history 
of galaxy clusters. This increases the proportion of early-type galaxies' 
members, rapidly building stellar mass (M$\star$) and cessation of SFR. 
A study of the origin of the dwarf elliptical in the Cluster Virgo using a 
simulation by Ref.\ \cite{boselli2008origin} concluded that one or more ram 
pressure stripping results in galaxies' starvation. The model predicts that 
all the star forming dwarf galaxies lose most of their gas contents when 
entering the cluster, which makes short the timescale for quenching of their 
star formation ($\leq 150$ Myr). Ref.\ \cite{kewley2006metallicity}, deriving 
the luminosity-metallicity relationship for local galaxy pairs, showed that 
members of galaxy pairs with smaller predicted distances possess lower 
metallicities, about the average of $\sim 0.2$ dex, than field galaxies of 
similar luminosity. Ref.\ \cite{ellison2008galaxy} found a lower metallicity 
offset of $\sim 0.1$ dex for the galaxy pair when compared with the field 
galaxies of similar luminosity. They also presented a comparison of 
metallicity-mass correlations between field mapped galaxies and again a low 
metallicity compensation of approximately $0.05$ dex was found. The minimized 
value of offset compared to metallicity-mass indicates that both higher as 
well as lower metallicity may influence the pairs to shift relative to 
the luminosity-metallicity relation. 

Several studies have tried to show the connection between interactions 
and AGN activities \cite{bickley2023agns,silva2021galaxy,gao2020mergers,
li2008interactions,schmitt2001frequency,croton2006many,bower2006breaking,
hopkins2005physical,cattaneo2005active,kang2005semianalytical,
schmitt2001frequency,croton2006many,bower2006breaking,hopkins2005physical,
cattaneo2005active,kang2005semianalytical}. Studies of the AGN fractions in 
pairs and field samples show that selecting galaxies with different contrast 
and anisotropy yields distinct AGN fractions, but the fractions for pairs and 
field samples are the same when a similar set of criteria is employed. This 
suggests that if AGN is detonated as a result of the interaction there is 
a possibility that this activity starts after the final stage, specifically 
once the merger has been finalised. When studying the relation of interaction 
with AGN activity, Ref.\ \cite{li2008interactions} observed that regardless 
of the statistics employed, the SFR increases for AGN with nearby neighbours 
as much as they do in inactive galaxies. This result contradicts the results 
from theoretical models that assume that black holes and AGN are directly 
linked with galaxy interactions and mergers \cite{schmitt2001frequency,
croton2006many,bower2006breaking,hopkins2005physical,cattaneo2005active,
kang2005semianalytical}. Observations from hydrodynamic simulations show 
that galaxy interactions may be responsible for the wind from the galaxy 
disk into the central region and cause star formation in the bulge to 
increase \cite{springel2000modelling,cox2006feedback,kapferer2005star,
meza2003simulations}. The study by Ref.\ \cite{kapferer2005star} found that 
galaxies metals increases when two or more galaxies undergo interactions, 
however other studies observed mergers to have lower metallicity when 
compared to non-mergers, a process which is linked with the high 
SFR \cite{sparre2022gas}. A study of dwarf galaxies by 
Ref.\ \cite{brosch2004interactions} concludes that galactic interactions can 
explain some of the central region patterns that drive star formation in 
large galaxies, but keeping in mind that it doesn't matter for dwarf 
galaxies. Ref.\ \cite{bergvall2003galaxy} found that interactions and mergers 
from star formation galaxies under the global aspect in general are not 
significantly different from field galaxies, therefore no change is expected 
with redshift. Authors also pointed out that galaxy formation is a long 
process and does not attain the maximum value at a specific redshift. 
Ref.\ \cite{smith2007spitzer}, using the Spitzer's mid-infrared study 
containing of 35 tidally distorted interacting galaxies which are in 
pre-merger stage shows that interactions from the tides of concentrate gas in 
the inner and central zones helps to fuel the galaxy which results in the 
rise of SFR believed to trigger the AGN.

Galaxy mergers and interactions are linked with morphology transformations, as 
it is believed from simulations that spiral galaxies can form ellipticals as 
they merge but it is not clear how this must happen, although it has 
been presented that disk galaxies can survive massive merger events 
\cite{toomre1972galactic,hopkins2008disks}. In this context, 
Ref.\ \cite{hopkins2008disks} states that ``the probability of survival of 
the disc depends on the properties of the progenitors such as the environment, 
gas content, and feedback mechanisms", which are related to the morphology of 
galaxies. Ref.\ \cite{van2008dependence}, using a sample of $4594$ galaxies 
obtained from the Sloan Digital Sky Survey (SDSS) Ref.\ \cite{york2000sloan} 
within the redshift 
between $0.02$ and $0.03$ and M$\star\geq 10^{10}$M$\odot$, found that 
morphology and structure have inherently different properties and depend 
differently on galaxies' stellar masses and galaxies' environments, whereas 
structure mainly depends on M$\star$ while morphology is environment 
dependent. Ref.\ \cite{park2007environmental} observed that variations of 
galaxy properties with environment originate from the morphology and 
luminosity dependence on environment due to the fact that galaxy properties 
depend on both luminosity and morphology that under a fixed luminosity and 
morphology, other galaxy properties are nearly independent with the 
environment. 

Recent studies have investigated merger environments. For example, 
Ref.\ \cite{kim2024distribution} carried out a study for current undergoing 
gravitational merging galaxies and having the sign of post-merger in clusters 
within the redshift ranges $0.05$ to $0.08$, and obtained that mergers are 
mostly found in low density environments, highlighting that it is likely 
post-mergers of galaxies merged in clusters' outskirts. The presence of 
sub-clusters within the rich clusters complicates this process, as it is 
clear that these 
structures can influence the merging process \cite{wen2024cataloguen}. Again, 
Ref.\ \cite{pearson2024effects} studying the effects of the galaxy environment 
on the merger fraction obtained that the merger fraction increases with the 
increase of local density and decreases with the increase of velocity 
dispersion. Ref.\ \cite{pearson2022north} generating the merging galaxies 
catalogue covering 5.4 square degrees around the North Ecliptic Pole (NEP) as 
detailed in Refs.\ \cite{toba2020search,ho2021photometric,
kim2021identification,oi2021subaru} with the redshift less than $0.3$, obtained 
that the merger fraction increases with redshift.  
Ref.\ \cite{sureshkumar2024galaxy}, determining whether mergers prefer denser 
or under dense environment using the Galaxy and Mass Assembly (GAMA) 
survey as detailed in Refs.\ \cite{driver2022galaxy,baldry2018galaxy,
driver2011galaxy,driver2009gama} with stellar mass less than $9.5$ M$\odot$, 
observed that the high velocity dispersion in dense environment prevent 
galaxy merging. Thus, the role played by the environment on merger rates and 
its properties like stellar mass, SFR, morphology, colours, chemical abundance 
are still in debate. 

Utilizing $3003$ galaxies of the Galaxy Zoo Project (GZP) from SDSS, 
Ref.\ \cite{darg2010galaxy} observed that mergers exhibit a wider range 
of colours and intense star formation, however, the distributions of the 
properties do not differ from non-mergers. They concluded that internal 
processes affect the mergers' detectability more than the external environment. 
Ref.\ \cite{casteels2014galaxy} analysing a galaxy sample from the GAMA survey, 
observed an increase in merger rates with galaxy masses and major mergers to 
be more common for massive galaxies, providing the conclusion that galaxy 
mergers depend on mass. On the other hand, recent studies revealed the 
variation of distributions and relationships of galaxy properties with the 
environment \cite{o2024effect,ding2024effects,privatus2025main,
privatus2025isolated,fernandez2024revealing}. 
Ref.\ \cite{fernandez2024revealing} observed a reduced star formation 
activity and older stellar population in rich groups when compared to 
poorer groups for ringed galaxies obtained from SDSS, showing the 
influence of group richness on ringed galaxy evolution. In this connection, 
analysing the distribution of merger and non-merger properties having 
different group richness will help to tackle the debated problem of 
whether group richness influences galaxy evolution.

Motivated by these findings, our study aims to investigate the influence 
of group richness on properties of distributions and relationships for 
mergers and non-mergers. Construction of samples based on group membership 
was retrieved from the catalogue by Ref.\ \cite{tempel2017merging}, where 
the authors carried out the study to determine galaxies' neighbours wherein the 
quantisation of group richness was obtained by employing the modified 
friends-of-friends algorithms on the twelfth release of the SDSS (DR 12) data. 
The mergers' sample was obtained from Ref.\ \cite{darg2010galaxy} selected from 
the GZP, a successful project which provide the visual observation of 
galaxies by majority of votes \cite{lintott2008galaxy}. 

The organization of the rest of this paper involves: Section \ref{secII}, 
where we explain the survey from which the sample was obtained, 
Section \ref{secIII} presents the methodology used to get the subsamples. 
Section \ref{secIV} presents the results, that is, distributions of galaxy 
properties and relationships, and Section \ref{secV} presents the discussion 
on the results. Finally, the summary and conclusion are provided in Section 
\ref{secVI}. We use the following parameters specified in 
Ref.~\cite{collaborartion2016planck}: Hubble constant
$H_0=67.8$ km s$^{-1}$ Mpc$^{-1}$, density of matter $\Omega_{m}=0.308$, 
and dark energy density $\Omega_{\Lambda} = 0.692$.

\section{Data} \label{secII}
As already stated, the galaxy samples in this study were obtained from the 
catalogue published by Ref.\ \cite{tempel2017merging}, constructed from the 
SDSS, specifically from the twelve data release (SDSS DR 12) 
\cite{eisenstein2011sdss,alam2015eleventh}. The methods of galaxy selection 
are as specified in Ref.\ \cite{tempel2017merging}, where the objects 
specified as GALAXY or QSO are selected, as also described well in 
Ref.\ \cite{alam2015eleventh} based on recommendations from the SDSS team. 
We applied a galactic extinction correction based on 
Ref.\ \cite{schlegel1998maps} criteria and removed fainter sources using the 
selection criterion as given by $m_r < 17.77$, where $m_r$ is the Petrosian 
apparent magnitude in the $r$-band. This magnitude limit reflects the 
incompleteness of the SDSS spectroscopic sample at fainter 
magnitudes \cite{strauss2002spectroscopic}. We corrected the redshift for the 
motion in relation to the cosmic microwave background (CMB) using the
equation as given by
\begin{equation}
	z_{\text{CMB}} = z_{\text{obs}} - v_p/c,
	\label{cmb}
\end{equation}
where $v_p$ is the movement along the line of sight relative to the CMB. The 
upper redshift limit was set to be $z = 0.1$, and the minimum redshift was 
set to be $0.03$ to ensure completeness, resulting in $8889$ galaxies.

We created the volume-limited sample from the flux-limited sample for 
uniformity due to the errors associated with the flux-limited sample, where at 
large distances faint objects are not observed \cite{teerikorpi2015eddington}. 
We transformed the apparent magnitude to absolute magnitude to increase 
the accuracy in distance measurements, especially for galaxies in groups due 
to their peculiar velocities, using the relation, 
 \begin{equation}
 	M_{r}= m_r-25-5\log_{10} (D_L)-K,
 	\label{eqR} 
 \end{equation}
where $D_L$ is the luminosity distance, $M_{r}$ is the absolute magnitude of 
$r$-band and $m_{r}$ is the apparent magnitude of $r$-band while $K$ is the 
combined correction in $k$ and evolution ($ev$). The calculation of 
$k$-correction was done using the KCORRECT ($v4\_2$) algorithm  
\cite{blanton2007k}, while estimation of evolution correction was done by 
using the evolution model for luminosity given by $k_{ev} = cz$, where 
$z = -1.62$ for the $r$-filter case \cite{blanton2003galaxy}. The estimation 
of magnitude and evolution correction at $z=0$ was done by assuming that 
luminosity is independent of distance \cite{tempel2014flux}. The effective 
maximum distance necessary to construct a luminous volume-limited sample was 
calculated using $m_{r,\text{lim}}=17.77$ mag, and $M_{r,\text{min}} =-\,20.55$
mag at which a total number of $8323$ galaxies were obtained.

\section{Methodology}
\label{secIII}
\subsection{Mass-Limited Samples}
\begin{figure}[h!]
	\subfigure{
		\includegraphics[width=0.33\linewidth]{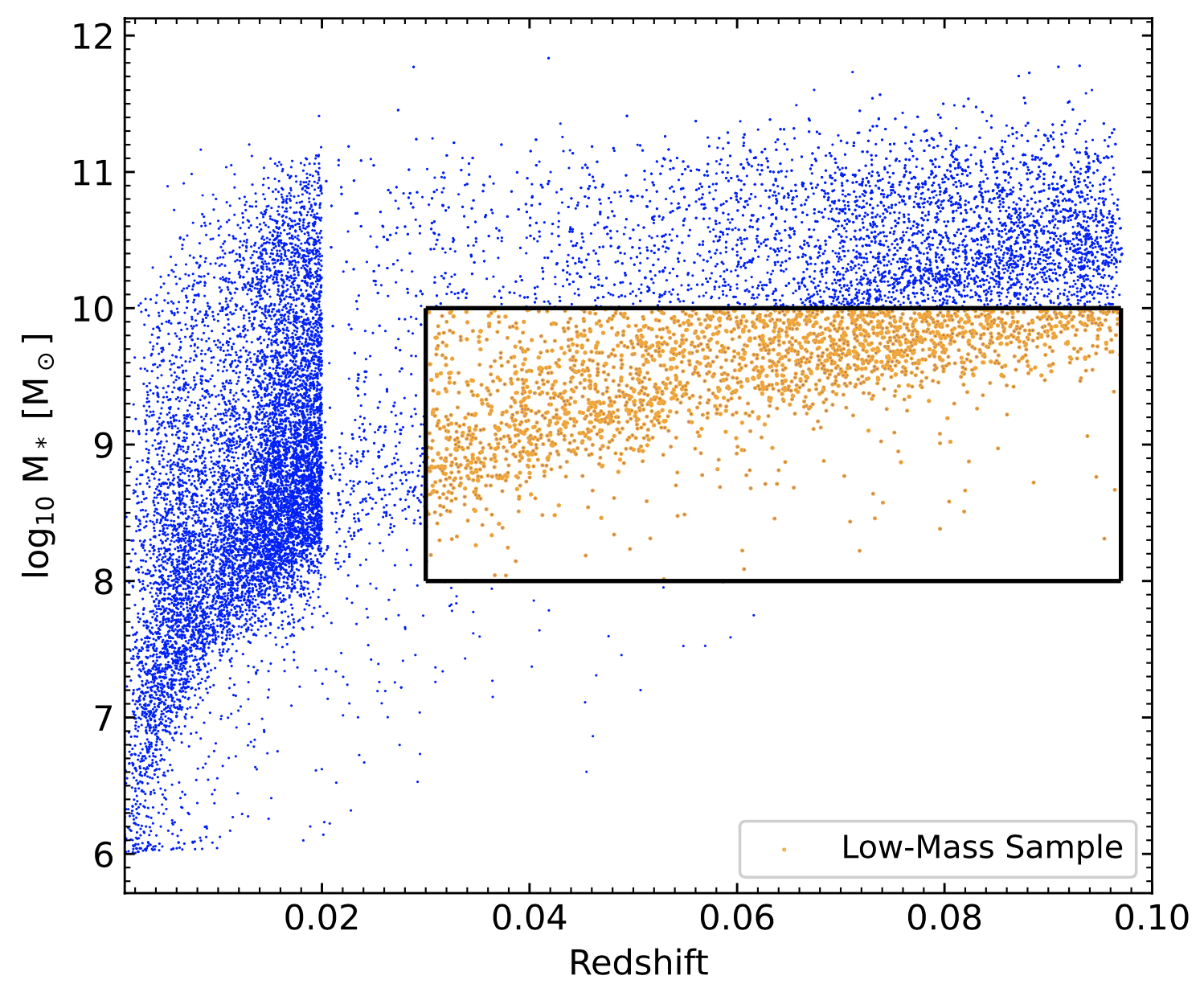}
		\label{MZL}
	}
	\hspace{-0.3cm}
	\subfigure{
		\includegraphics[width=0.33\linewidth]{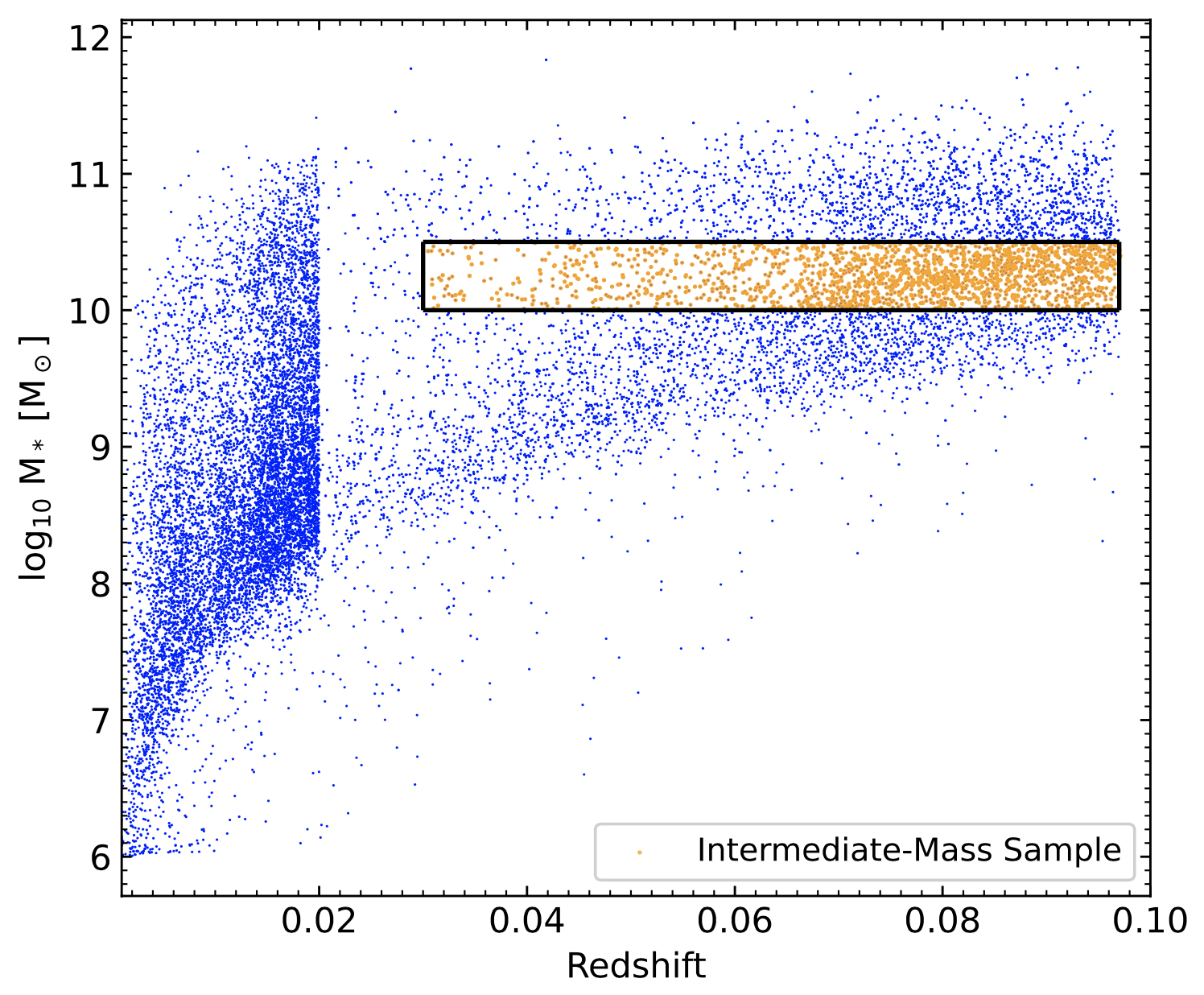}
		\label{MZI}
	}
	\hspace{-0.3cm}
	\subfigure{
		\includegraphics[width=0.33\linewidth]{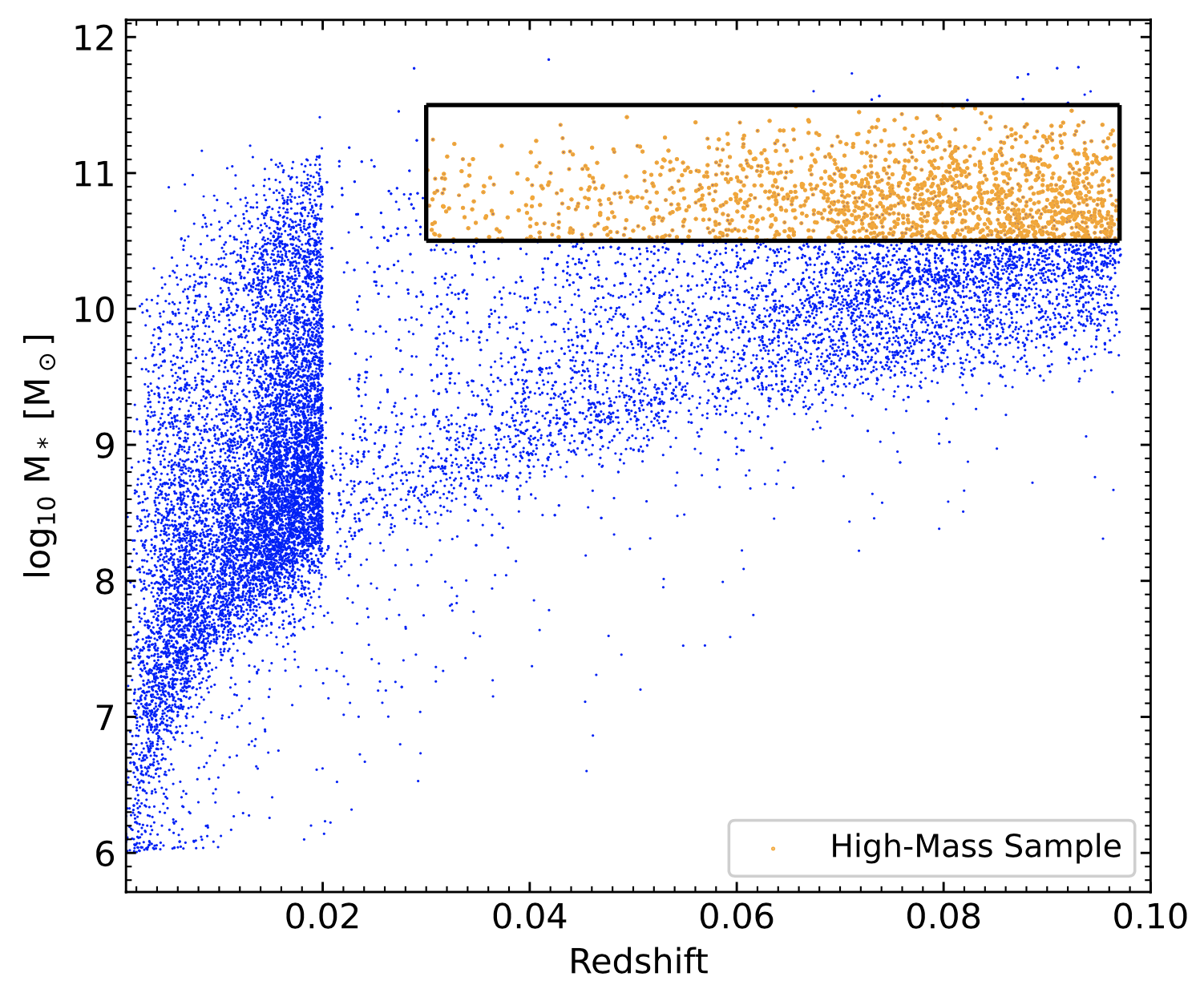}
		\label{MZH}
	}
	\vspace{-0.4cm}
 	\caption{Redshift-stellar mass diagrams for the subsamples of galaxies
out of 8889 galaxies. The orange dots inside the rectangle show the low-mass, 
intermediate-mass, and high-mass (from left to right) galaxies.}
 	\label{MZ}
 \end{figure}

Using the obtained sample as detailed in Section \ref{secII}, we further 
constructed the mass-limited subsamples to investigate the influence of 
mass as it is well known that when the galaxies have a significant mass 
difference the observed changes in other properties may be mass related. 
The stellar masses used in this study are those calculated by the Max Planck 
Institute for Astrophysics and Johns Hopkins University (MPA\&JHU), 
in which they employed the Bayesian approach (refer to 
Ref.\ \cite{kauffmann2004environmental}) where the calculation of 
spectroscopic fibre aperture depends on the fibre magnitudes while total 
M$\star$ relies on model magnitudes. The low-mass 
($8\leq \log_{10} $M$\star< 10$), intermediate-mass 
($10\leq \log_{10} $M$\star< 10.5$), and high-mass 
($10.5\leq \log_{10} $M$\star< 11.5$) samples contain $3051$, $2439$, 
and $2833$ galaxies, respectively as shown in Fig.~\ref{MZ}.
 
 \subsection{Constructing Mergers and Non-Mergers} \label{mn}
\begin{figure}[h!]
	\subfigure{
		\includegraphics[width=0.33\linewidth]{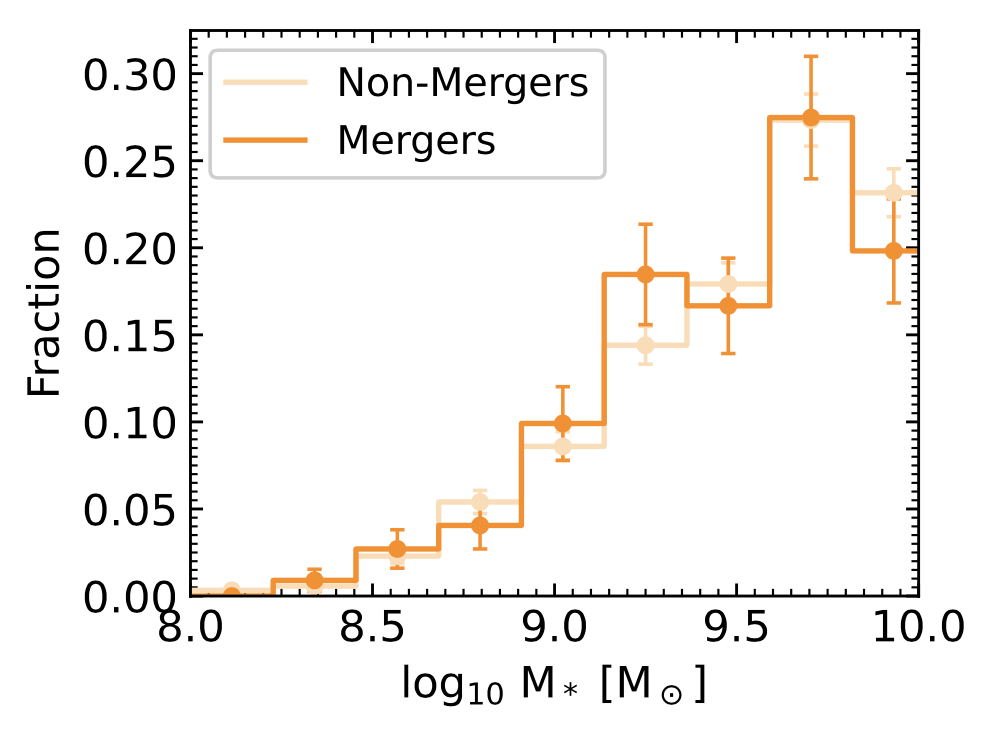}
	}
	\hspace{-0.3cm}
	\subfigure{
		\includegraphics[width=0.33\linewidth]{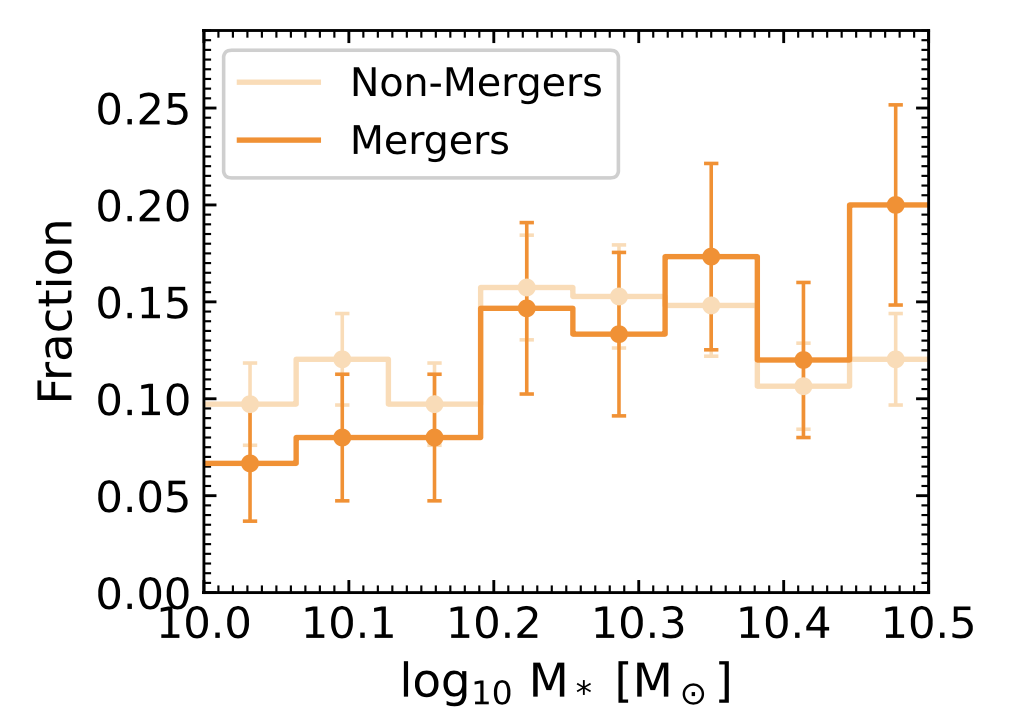}
	}
	\hspace{-0.3cm}
	\subfigure{
		\includegraphics[width=0.33\linewidth]{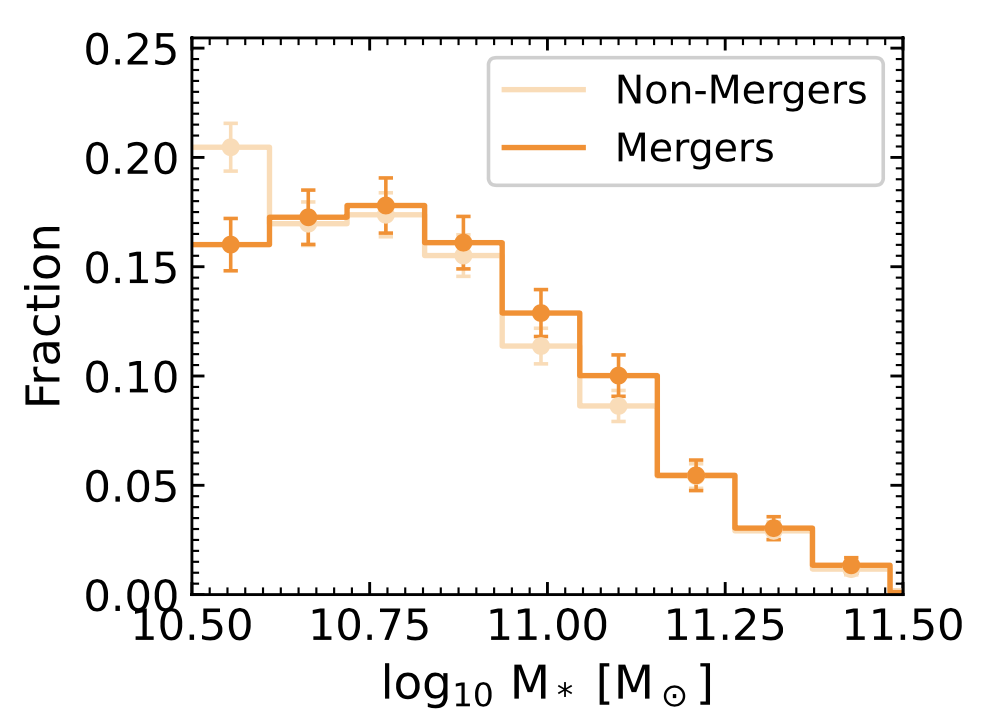}
	}
	\hspace{-0.3cm}
	\subfigure{
		\includegraphics[width=0.33\linewidth]{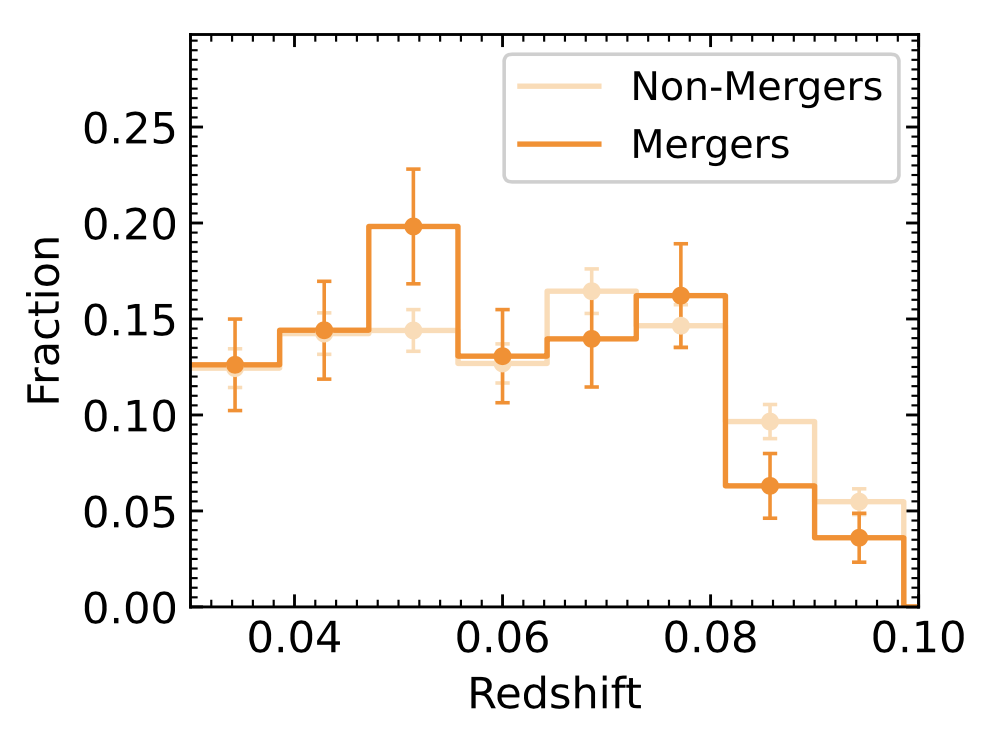}
	}
	\hspace{-0.3cm}
	\subfigure{
		\includegraphics[width=0.33\linewidth]{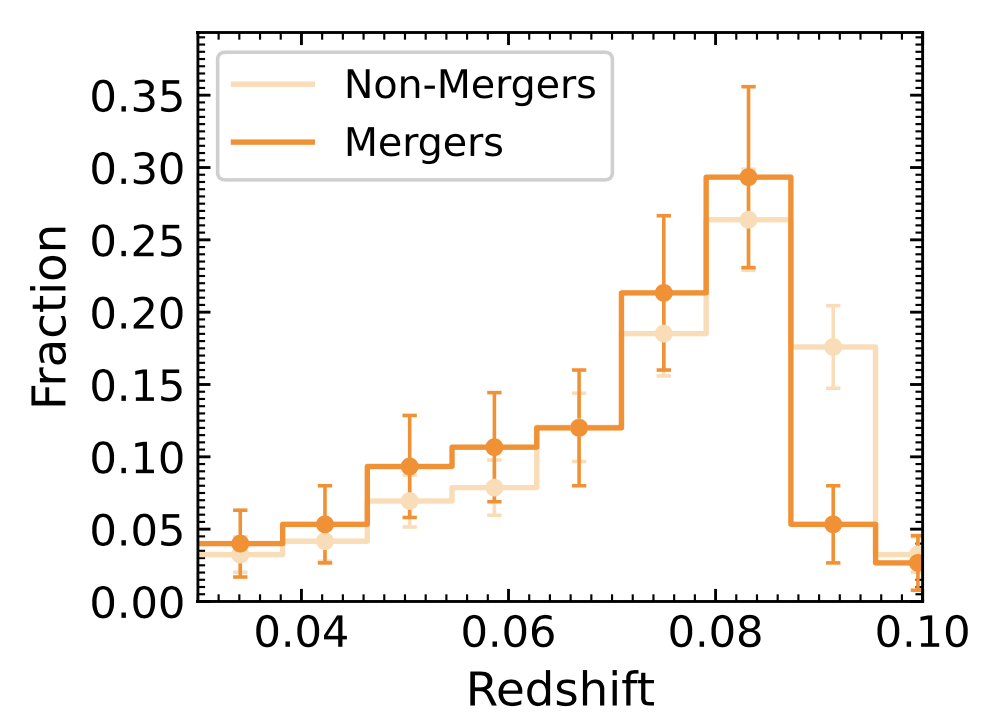}
	}
	\hspace{-0.3cm}
	\subfigure{
		\includegraphics[width=0.33\linewidth]{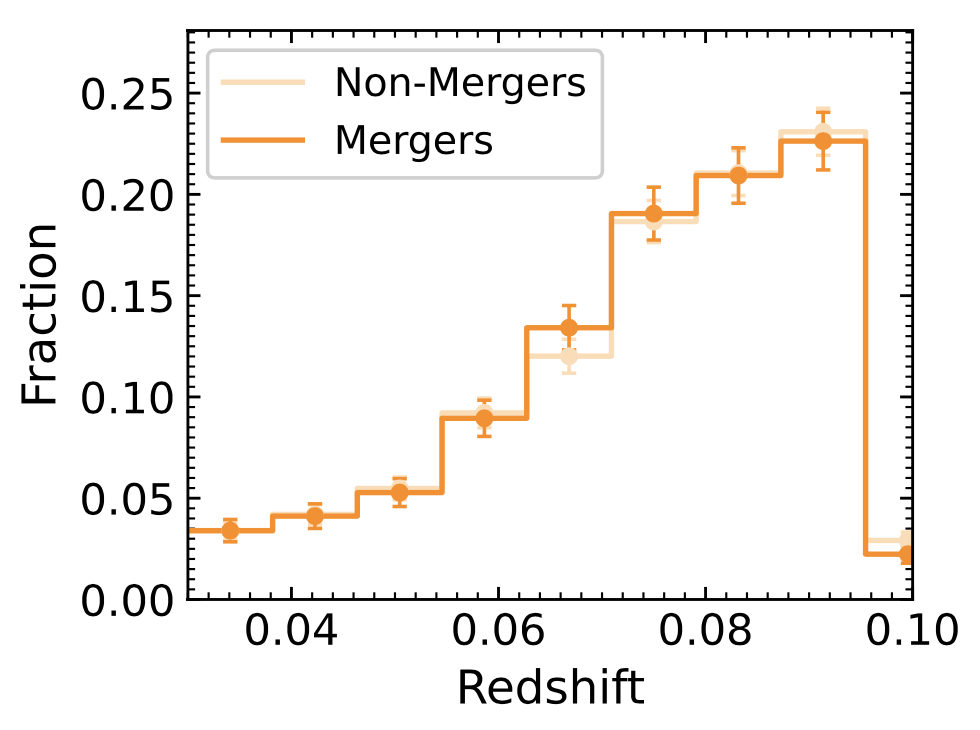}
	}
 	\vspace{-0.4cm}
 	\caption{Stellar mass (top panels) and redshift (bottom panels) 
 		distributions of mergers (dark) and non-mergers (light) for 
low-mass, intermediate-mass, and high-mass (from left to right, respectively) 
subsamples. The error bars are 1 $\sigma$ Poissonian errors.}
 	\label{MZD}
 \end{figure}

As already stated, the merging pairs were obtained from 
Ref.\ \cite{darg2010galaxy} constructed from the GZP catalogue 
\cite{lintott2008galaxy}, whose complete description, including the project 
and preliminary data reduction, is detailed in 
Refs.\ \cite{lintott2008galaxy}. GZP is a successful project that collectively 
offered classifications for $\sim 900000$ images from SDSS, used in a 
number of publications \cite{land2008galaxy,slosar2009galaxy,bamford2009galaxy,
skibba2009galaxy,schawinski2009galaxy,lintott2009galaxy,cardamone2009galaxy,
lintott2011galaxy}. In this project, users were asked to classify galaxies, 
including whether there are mergers (M). The catalogue was built by using 
weighted-merger-vote fraction ($f_M$) given by Eq.~\eqref{fm} which was 
obtained by dividing the number of mergers classification $n_M$ to the 
total number of classification $n_{E, S, B, M}$, where $E$, $S$ and $B$
are elliptical, spiral and bad image of star, respectively for a particular 
object multiplied by a specific weighting factor $W$ that measures the quality 
of users that have provided the classification for the object 
\cite{lintott2008galaxy}.  
 \begin{equation}
 	f_M =\frac{Wn_M}{n_{E, S, B, M}}.
 	\label{fm}
 \end{equation}
$f_M$ ranges $0\leq f_M \leq 1 $, where $1$ implies ``unmistakably a merger" 
while $0$ indicates ``nothing like a merger". The galaxies with 
$0.4\leq f_M \leq 1 $ were considered as mergers. A total number of $470$ 
low-mass, $526$ intermediate-mass and $1118$ high-mass mergers were 
obtained. Similarly for non-mergers, a total number of $2581$ low-mass, 
$1913$ intermediate-mass and $1715$ high-mass galaxies were found. The 
distributions of stellar masses and redshifts for mergers and non-mergers 
samples are shown in Fig.~\ref{MZD}. We performed the 
Kolmogorov–Smirnov (KS) statistical test detailed in Refs.\
 \cite{hodges1958significance,harari2009kolmogorov} to compare the 
distributions of stellar masses and redshifts of mergers and non-mergers. For 
the low-mass sample, the 
KS statistics and corresponding p-values (in brackets) were $0.05$ ($0.715$)
for stellar mass, and $0.07$ ($0.268$) for redshift. For the intermediate-mass 
sample, the results were $0.13$ ($0.480$) for stellar mass, and 
$0.15$ ($0.147)$ for redshift. Finally, for the high-mass sample, the values 
were $0.05$ ($0.079$) for stellar mass, and $0.02$ ($0.931$) for redshift. 
Since the KS statistics are closer to zero and the p-values are greater than 
the standard value ($0.05$) for all the distributions of stellar masses and 
redshift in the three subsamples, this implies that the mergers and non-mergers 
originate from the same population. Hence, the following analysis is
neither influenced by stellar mass nor redshift, supporting the observations 
from the histogram figures.

\subsection{Quantification of Group Richness}\label{qn}
With the aim of comparing the distributions of mergers' sample (SM) and 
non-mergers' sample (SNM) with different group richness, the membership 
details provided in Ref.\ \cite{tempel2017merging} were used, in which 
the galaxies are assigned into groups using modified friends-of-friends 
algorithms (FoF) \cite{tago2008groups,tago2010groups}. This technique involves
using a specific radius, sometimes referred to as linking length, which varies 
with redshift but should be minimum to make sure that the properties are 
invariant with distance \cite{tempel2017merging}. This linking length is 
given by an arctan function as 
\begin{equation}
	d_{LL}(z) = d_{LL,0} \left[1 + a \arctan\left(\frac{z}{z_*}\right)\right],
	\label{arc} 
\end{equation}
where $d_{LL,0}$ implies the radius at initial redshift, $a$ and ${z_*}$ are 
free parameters. $d_{LL}(z)$ is the linking length in which the group of 
galaxies is defined at a specific redshift $z$. The values of $d_{LL,0}$ is 
$0.34$ Mpc, $a$ is $1.4$ and ${z_*}$ is $0.09$, which were obtained by fitting 
Eq.~\eqref{arc} using observational data. The mergers' (S1M, S2M, S3M) and 
non-merges' (S1NM, S2NM, S3NM) subsamples were constructed based on the 
number of galaxies in the group where they exist ($N_{rich}$). Here, S1 
($N_{rich} = 1$), S2 ($2\leq N_{rich} \leq 5$) and S3 
($6\leq N_{rich} \leq 50$) indicate the poor to rich group systems, 
respectively following the definition by Ref.\ \cite{tempel2017merging} 
that $N_{rich} \geq 6$ are more reliable group. For mergers, a total number of 
$222\,(47\%)$, $187\,(40\%)$, $61\,(13\%)$ for low-mass; 
$191\,(36\%)$, $253\,(48\%)$, $82\,(16\%)$ for intermediate-mass; 
$269\,(24\%)$, $561\,(50\%)$, $288\,(26\%)$ for high-mass representing S1M, 
S2M and S3M respectively were obtained. For non-mergers a total number 
of $1222\,(47\%)$, $1005\,(40\%)$, $354\,(14\%)$ for low-mass; 
$801\,(42\%)$, $788\,(41\%)$, $324\,(17\%)$ for intermediate-mass; 
$508\,(30\%)$, $817\,(48\%)$, $390\,(23\%)$ for high-mass 
representing S1NM, S2NM and S3NM respectively were obtained. 
These volume-limited and mass-limited samples are used throughout the study 
unless we state otherwise.

\section{results}\label{secIV}
The results for distributions and relationships of the properties between 
mergers and non-mergers, and also the influence of group richness are analysed 
in this section using the samples obtained in Section \ref{qn}. The 
distribution of SFR, specific SFR (SSFR), D$_n$ (4000) and $u-r$ colour, and
their variations with stellar mass are analysed. As already stated, the 
stellar mass was obtained by the Bayesian approach 
\cite{kauffmann2004environmental}, where the spectroscopic fibre aperture 
calculation depends on the fibre magnitudes while total M$\star$ relies on 
model magnitudes. 
The total SFR was estimated using the methods described in 
Refs.\ \cite{brinchmann2004physical,tremonti2004origin} by the MPA\&JHU team 
for star forming (SF) galaxies while making adjustments for galaxies that are 
non-star forming (non-SF). The H$\alpha$ calibration, as well explained 
in Ref.\ \cite{kennicutt1998star}, was used to calculate the SFR for SF. 
Keeping a note that the description by Ref.\ \cite{brinchmann2004physical} 
involves SFR calculation within the galaxy fiber aperture. The methods 
outlined in Ref.\ \cite{salim2015mass} were used to estimate the SFR beyond 
that region. For AGNs and galaxies with weak emission lines, the 
SFR was estimated while considering that ionization in 
non-star forming galaxies may originate from sources such as 
post-AGB stars or AGN activity. Therefore, the SFR derived from H$\alpha$ 
should be regarded as an upper limit \cite{sanchez2018sdss}. The spectral index 
D$_n$ (4000) is used as the tracer of the age of stellar population, 
calculated due to the spectral difference which occurs at $4000 \text{\AA}$, 
because of the accumulation of many spectral lines in a limited region 
\cite{kauffmann2003host}. The definition D$_n$ (4000) as detailed in 
Ref.\ \cite{balogh1999differential}, given as the ratio of average flux density 
in a narrow continuous band ($3850 - 3950$ $\text{\AA}$ and 
$4000-4100$ $\text{\AA}$) is used in this study. 
\subsection{Distributions of Properties} 
\begin{figure}[h!]
	\subfigure{
		\includegraphics[width=0.32\linewidth]{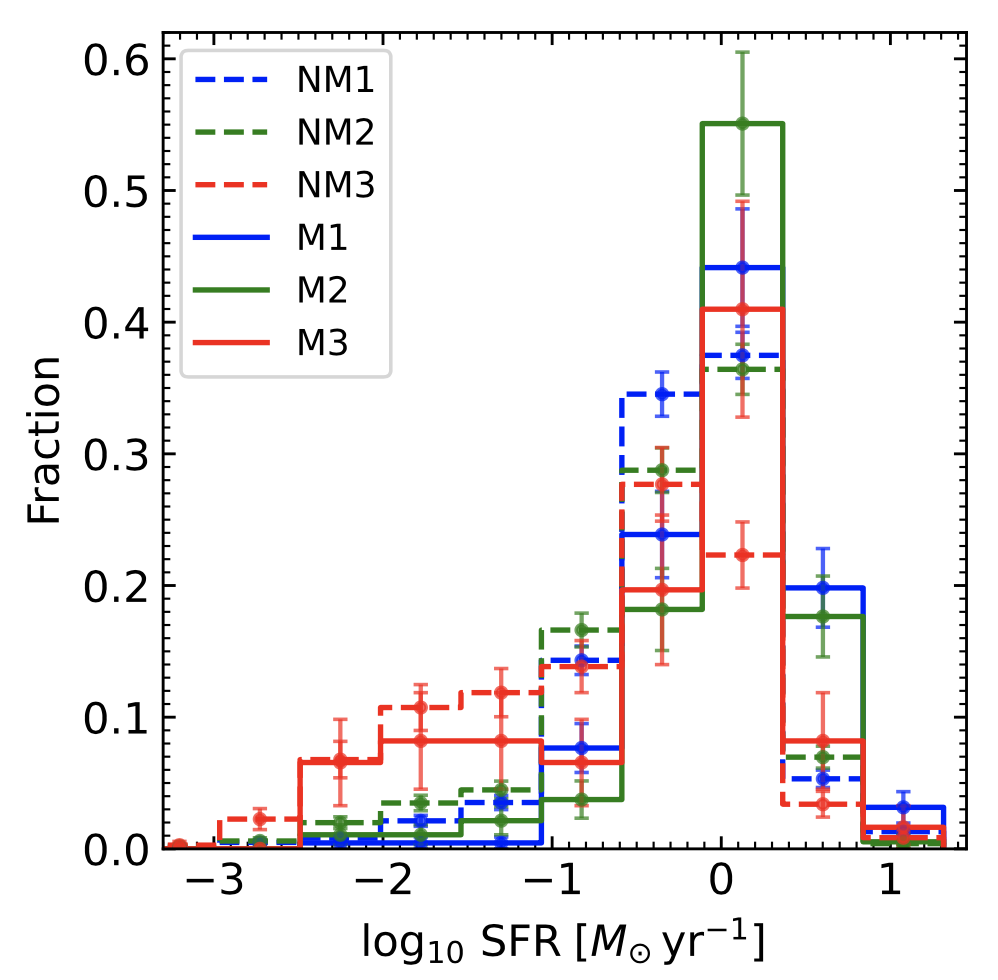}
		\label{SFL}
	}
	\subfigure{
		\includegraphics[width=0.32\linewidth]{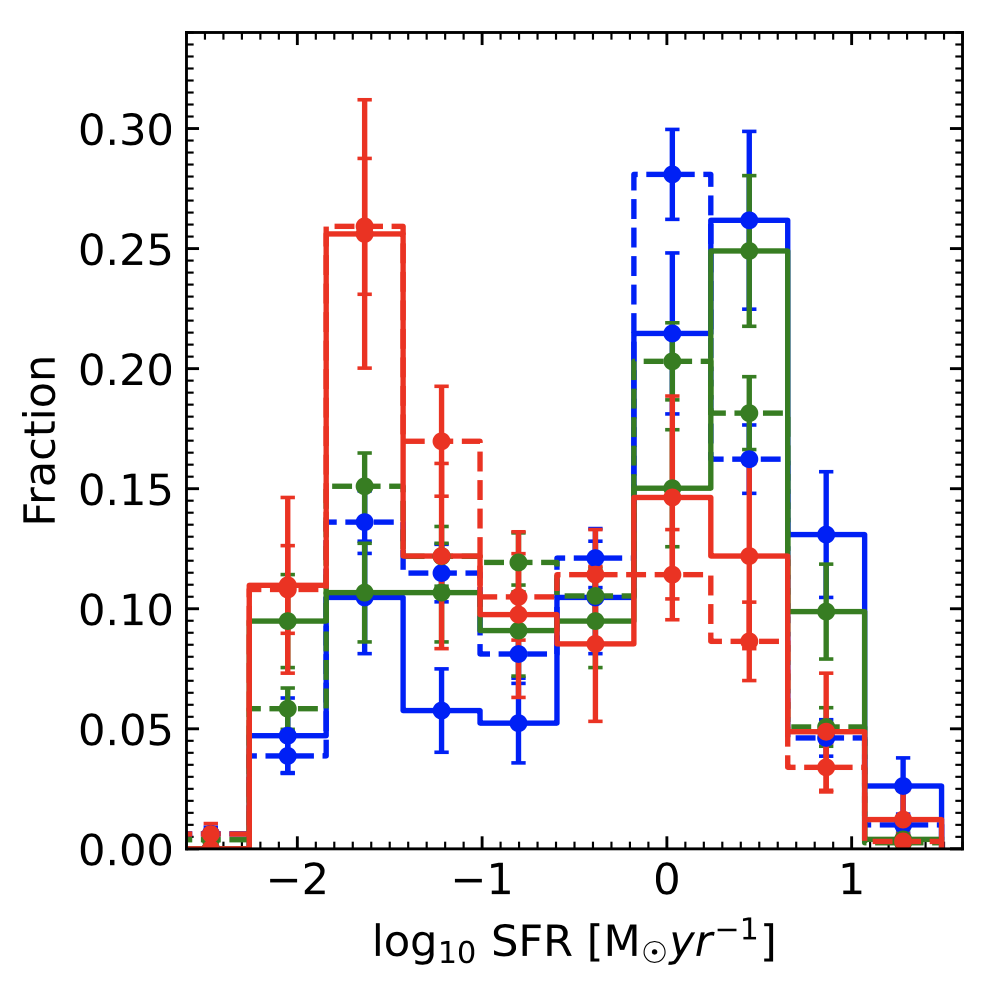}
		\label{SFI}
	}
	\subfigure{
		\includegraphics[width=0.32\linewidth]{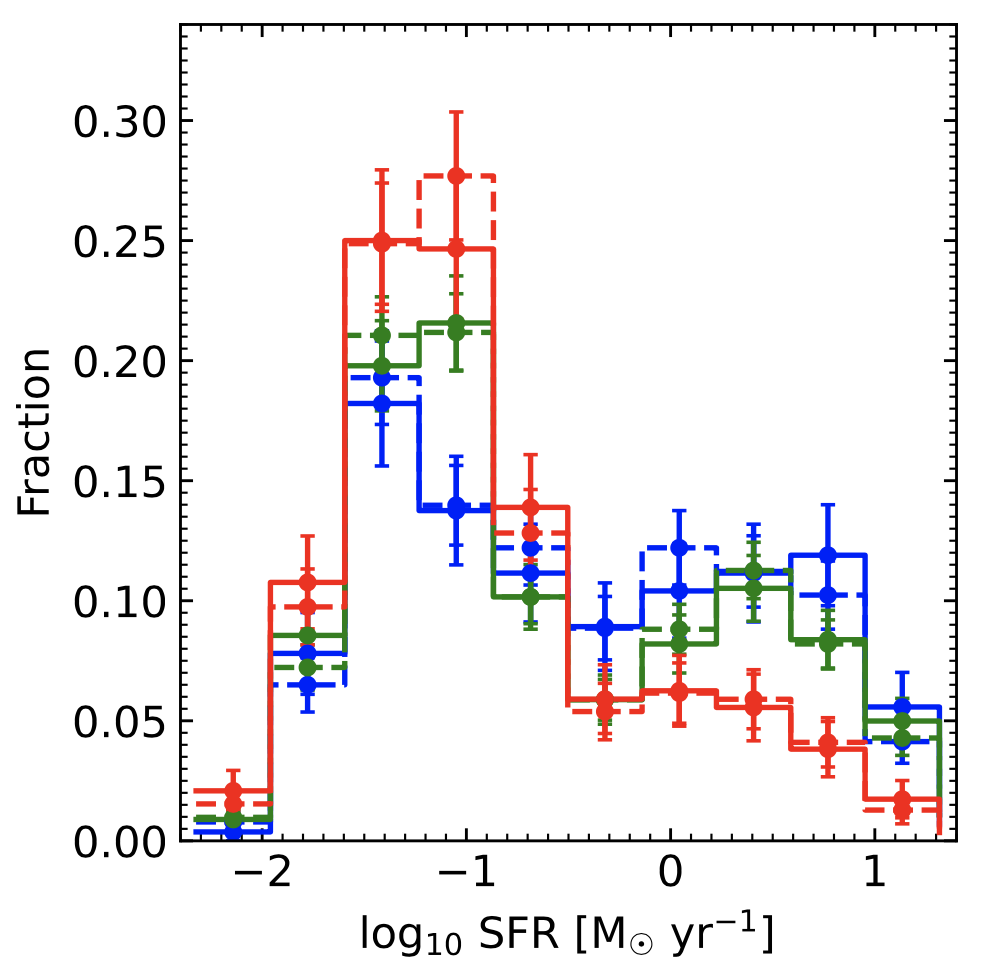}
		\label{SFH}
	}
		\subfigure{
		\includegraphics[width=0.32\linewidth]{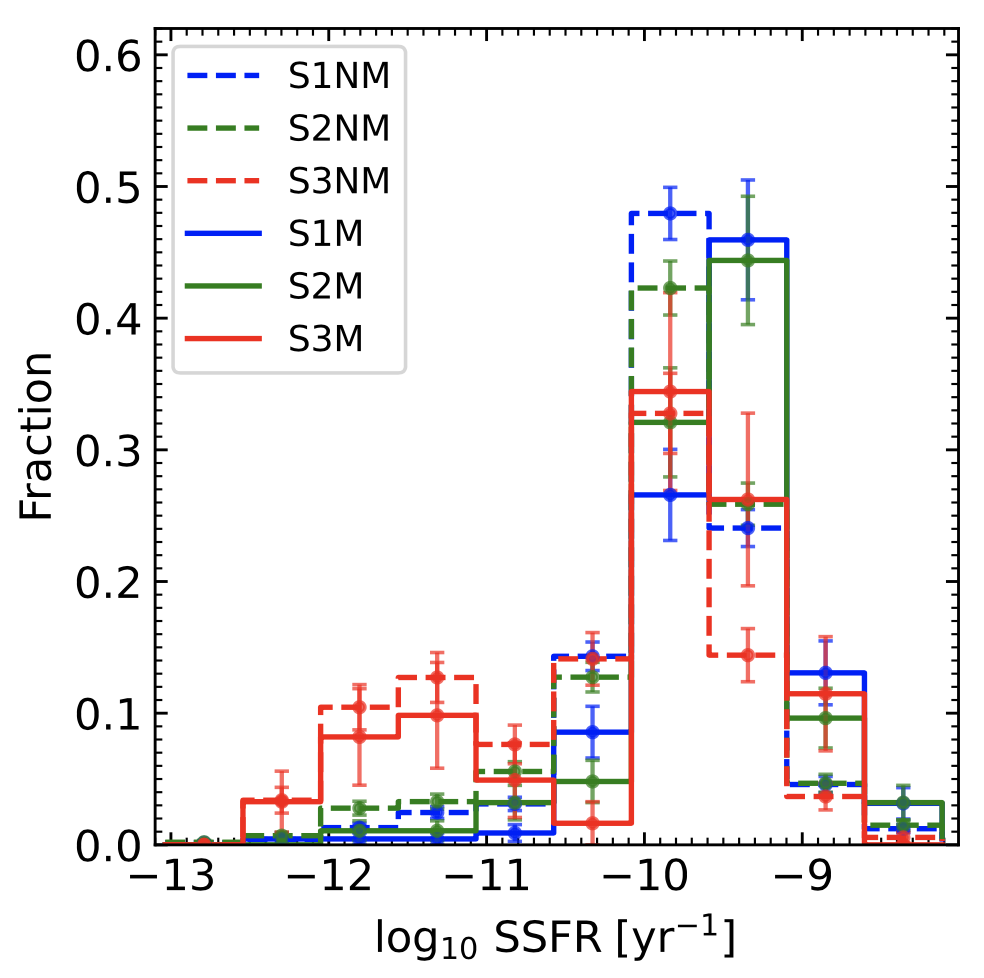}
		\label{SSFL}
	}
	\subfigure{
		\includegraphics[width=0.32\linewidth]{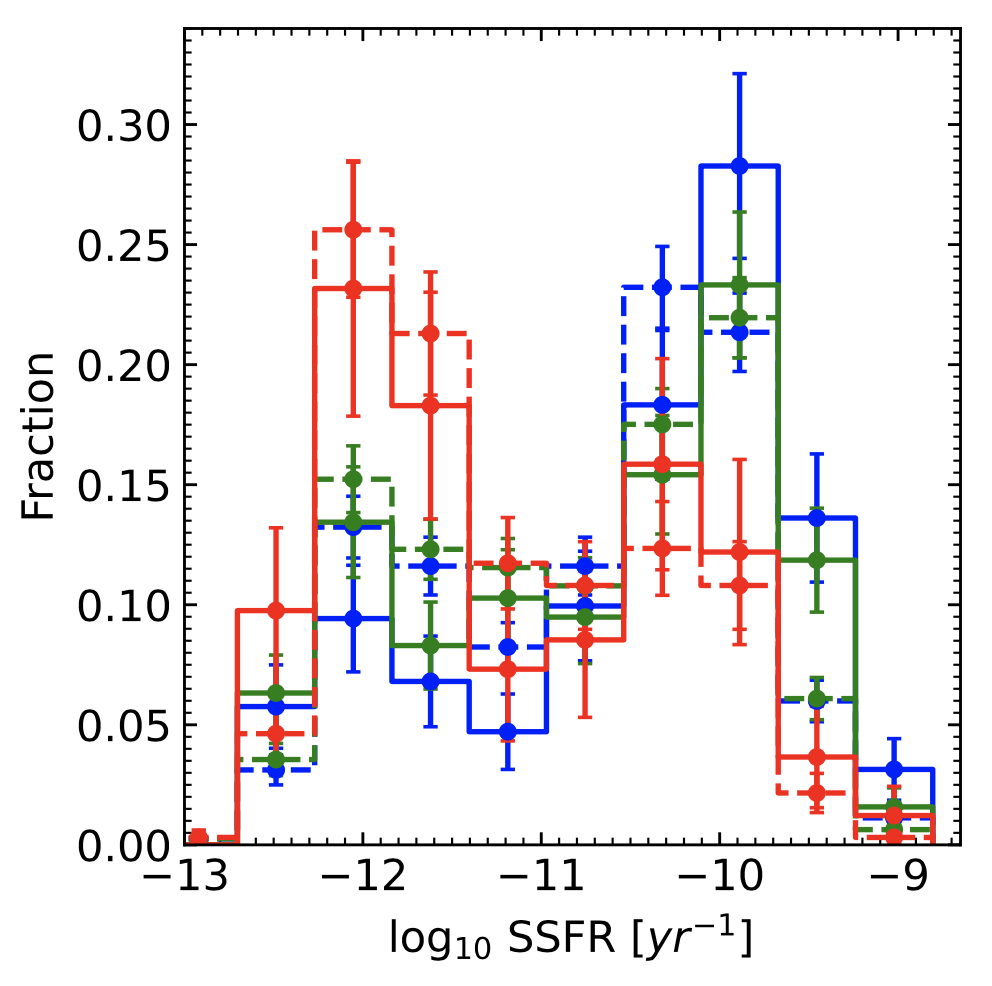}
		\label{SSFI}
	}
	\subfigure{
		\includegraphics[width=0.32\linewidth]{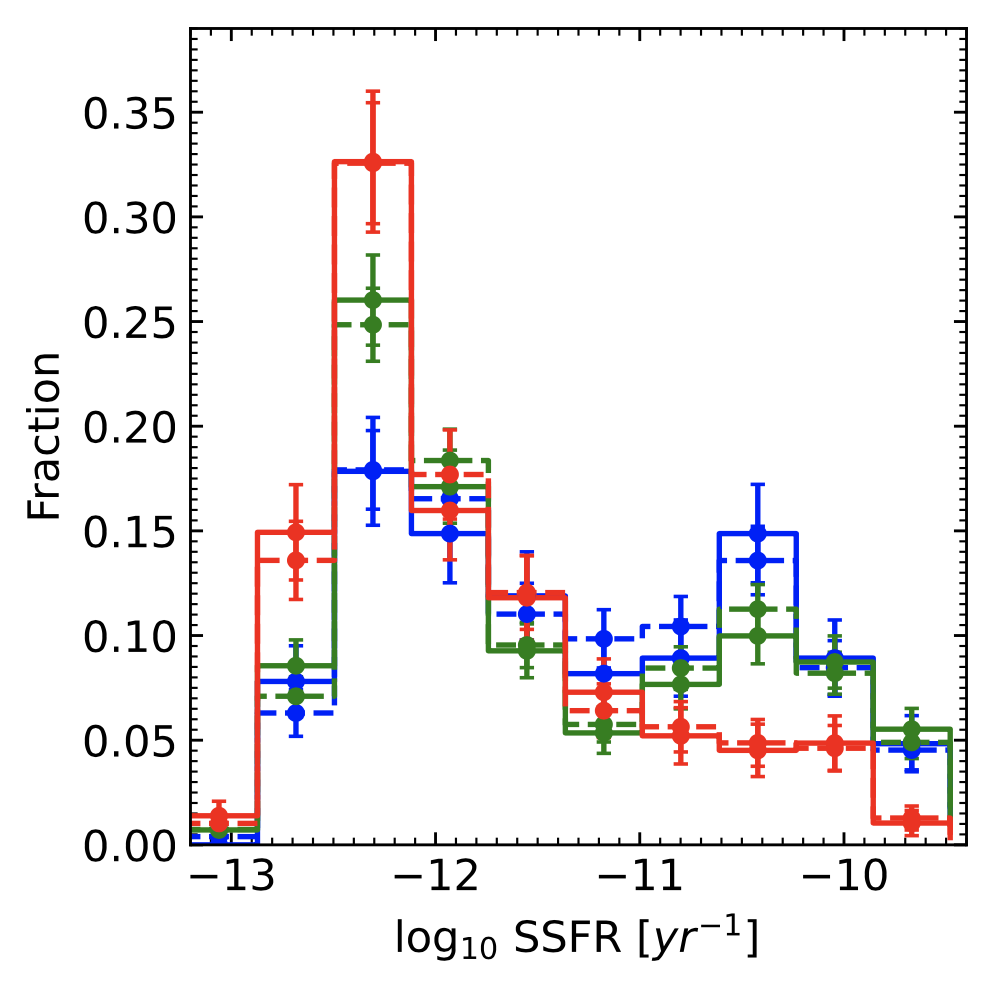}
		\label{SSFH}
	}
	\vspace{-0.2cm}
\caption{Star formation rate (top panels) and specific star formation 
		rate (bottom panels) distributions of mergers (solid lines) and non-mergers 
		(dashed line) for low-mass, intermediate-mass, and high-mass (from left to 
		right, respectively) subsamples. The error bars in this figure and all other 
		similar figures are $1\sigma$ Poissonian errors. 
		S1NM, S2NM, S3NM are non-mergers 
		while S1M, S2M, S3M are mergers sub-samples arranged from 
		poor to rich group system, respectively.}
	\label{SFS}
\end{figure}

\begin{figure}[h!]
	\subfigure{
		\includegraphics[width=0.32\linewidth]{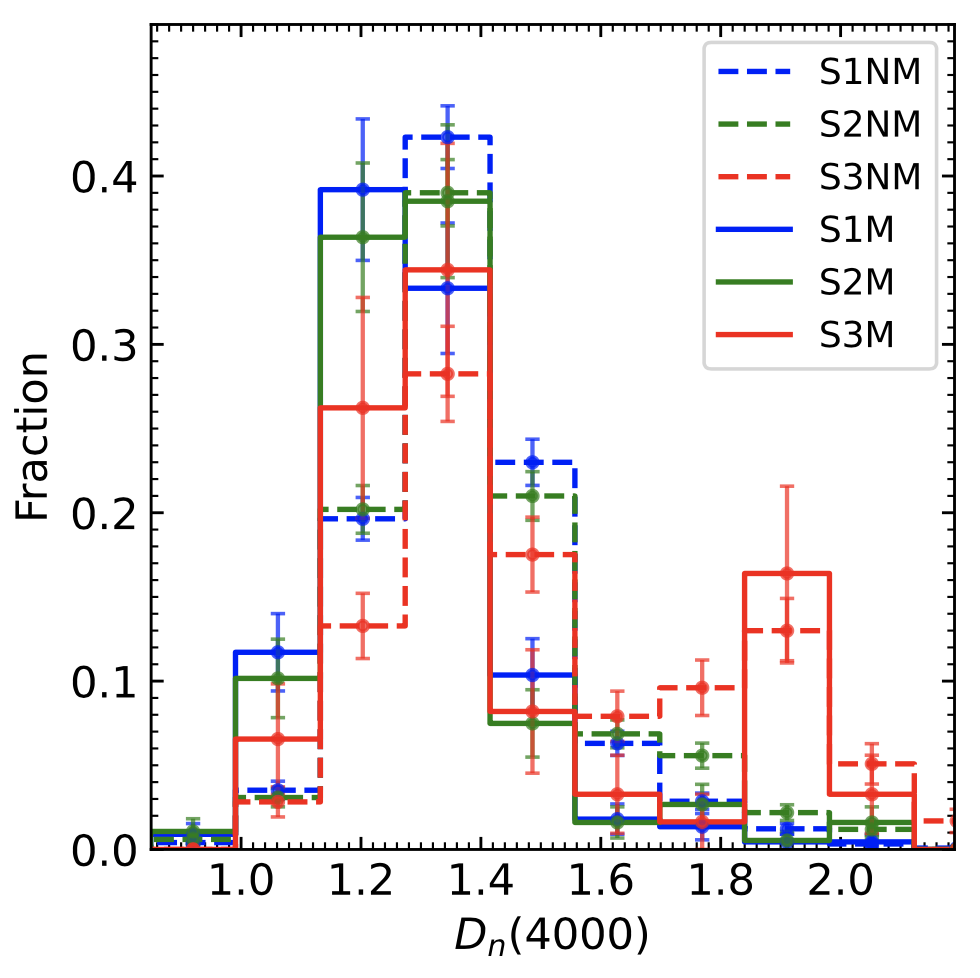}
		\label{DL}
	}
	\subfigure{
		\includegraphics[width=0.32\linewidth]{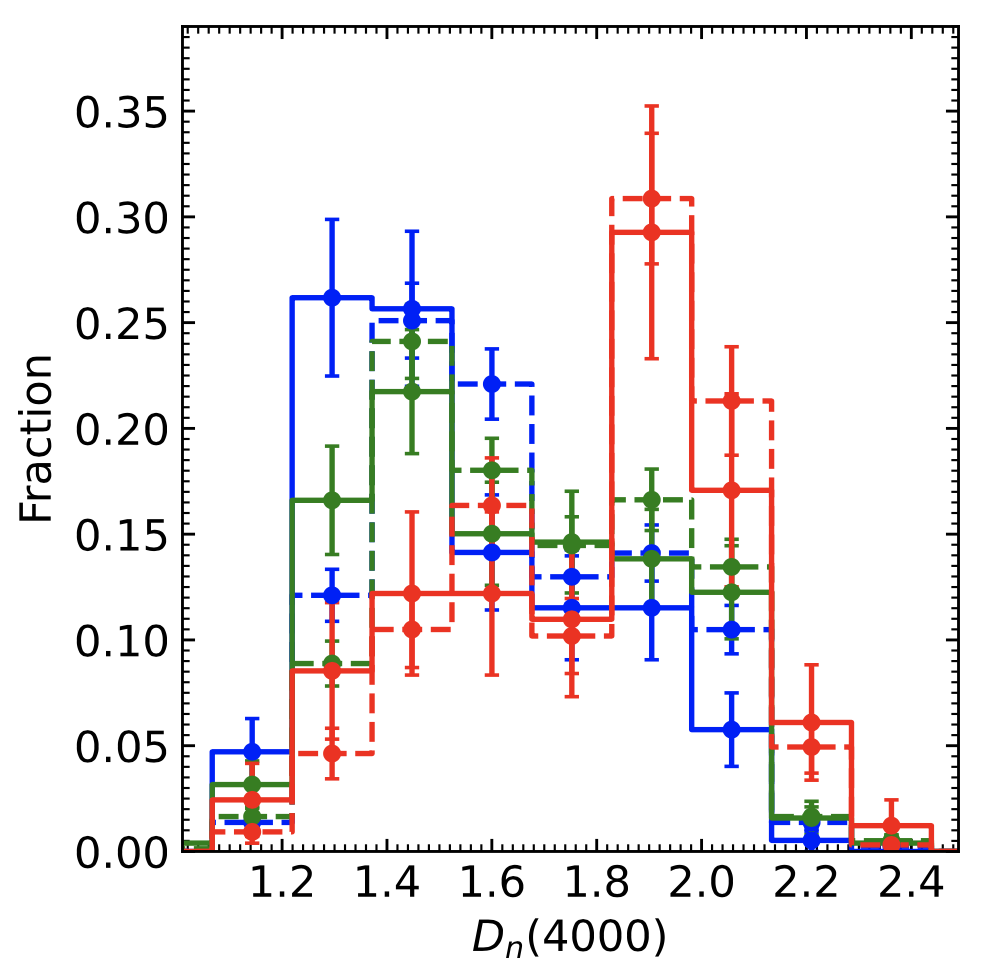}
		\label{DI}
	}
	\subfigure{
		\includegraphics[width=0.32\linewidth]{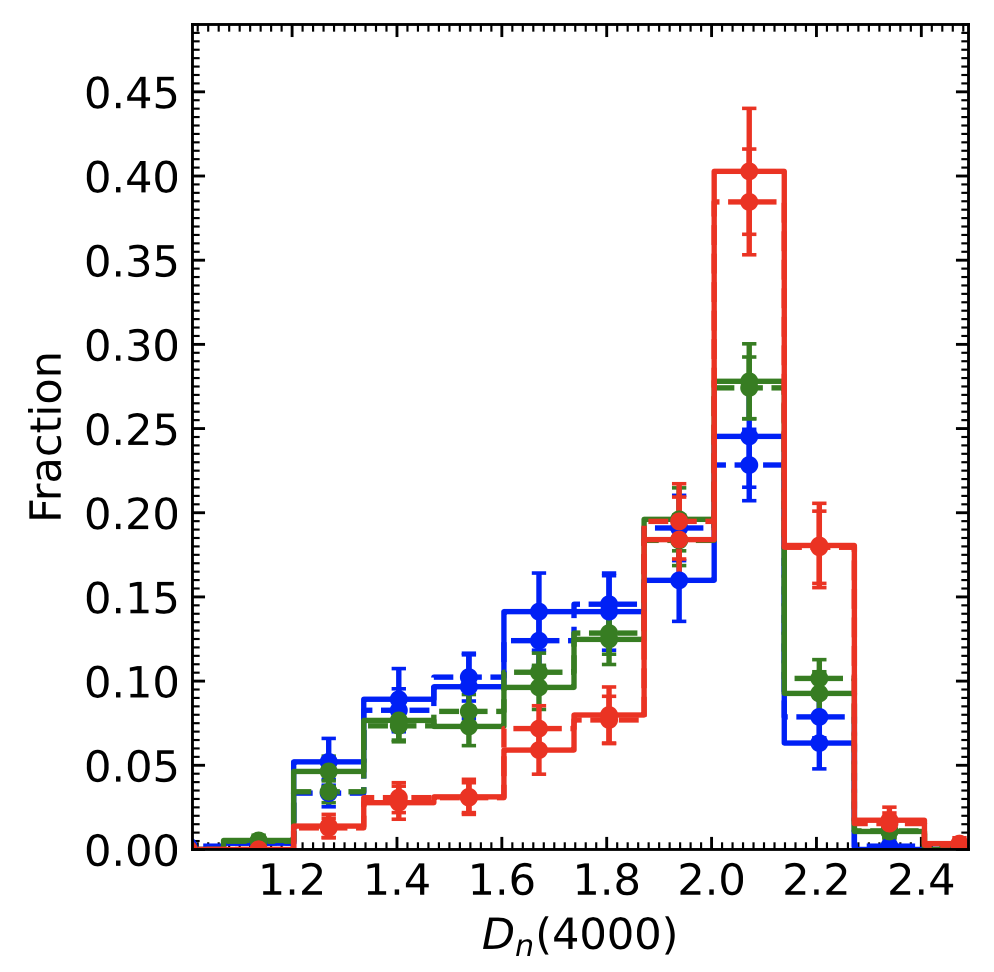}
		\label{DH}
	}
		\subfigure{
		\includegraphics[width=0.32\linewidth]{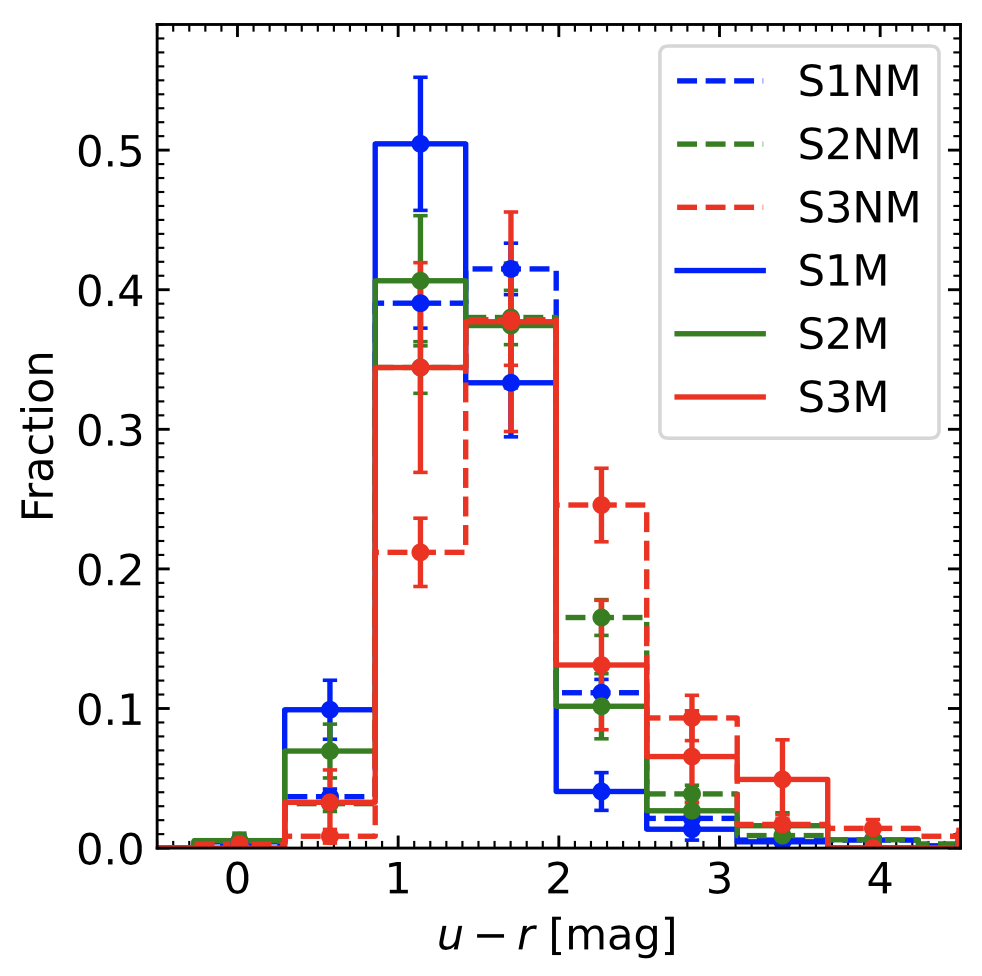}
		\label{CL}
	}
	\subfigure{
		\includegraphics[width=0.32\linewidth]{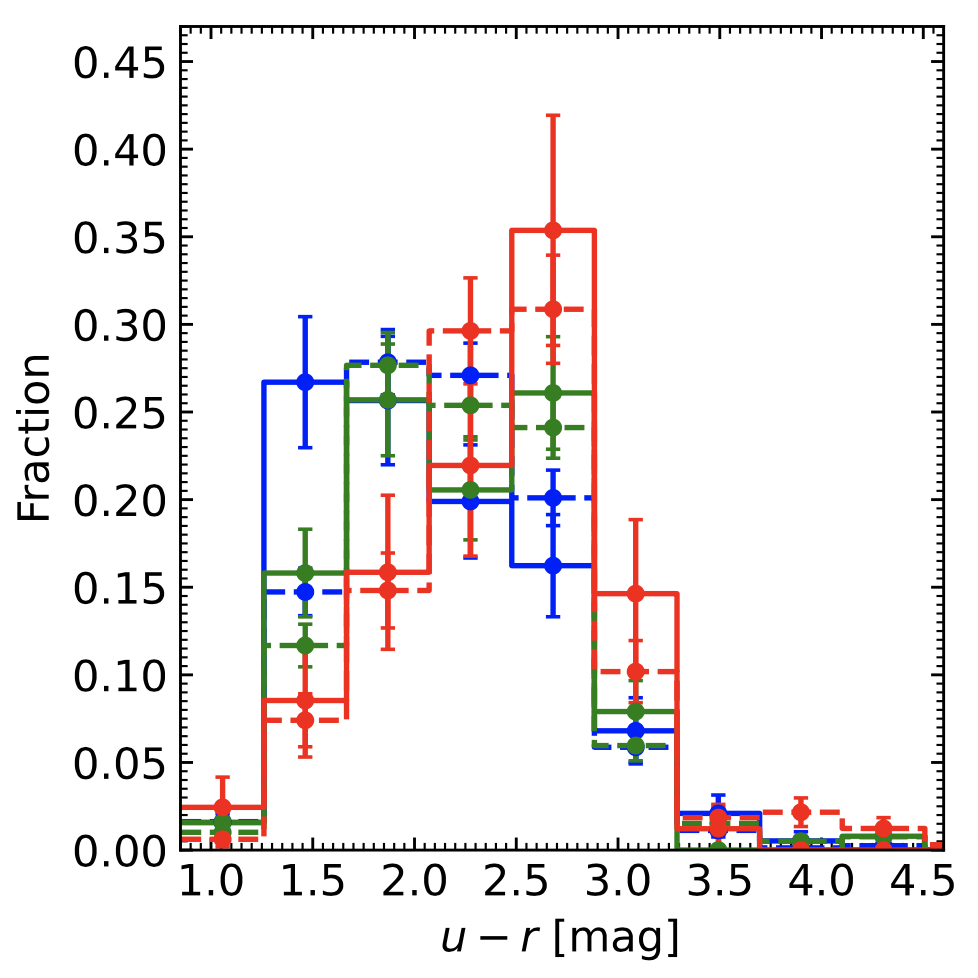}
		\label{CI}
	}
	\subfigure{
		\includegraphics[width=0.32\linewidth]{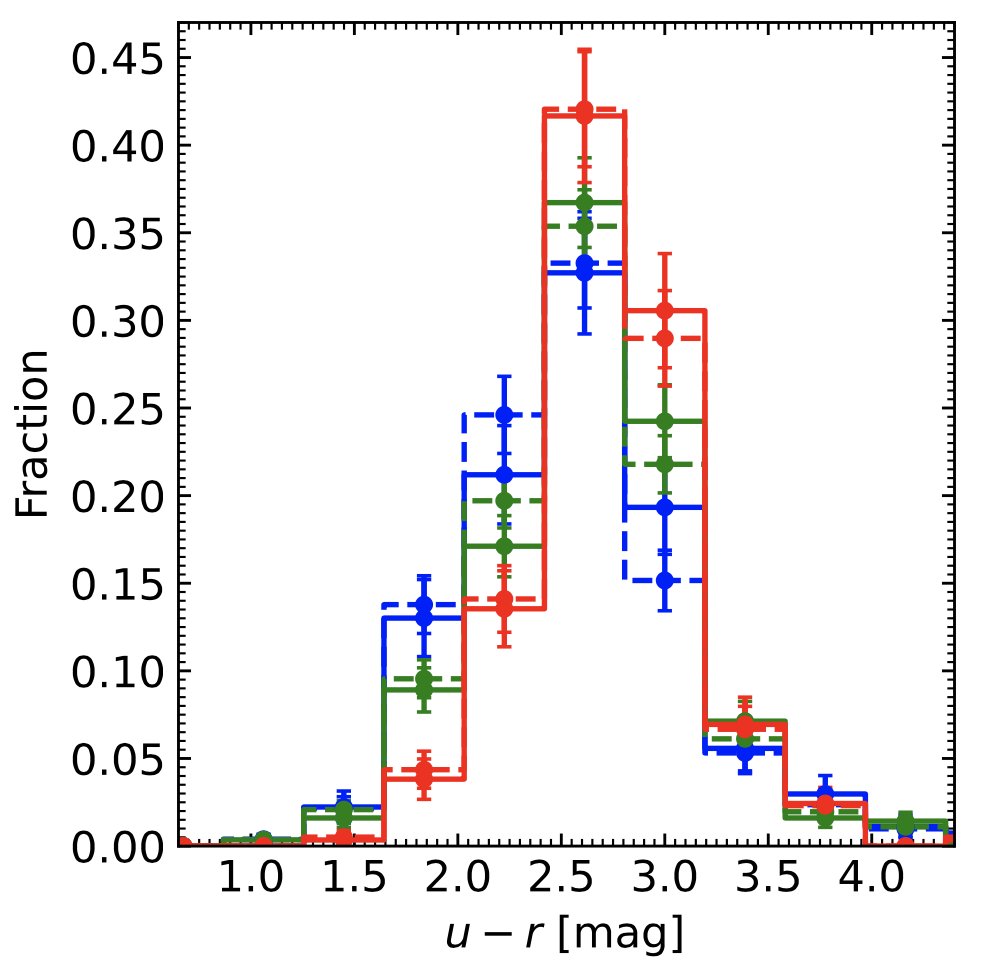}
		\label{CH}
	}
	\vspace{-0.2cm}
	\caption{D$_n$ (4000) (top panels) and $u-r$ colour (bottom panels) 
distributions of non-mergers (dashed lines) and mergers (solid lines) for 
low-mass, intermediate-mass, and high-mass (from left to right, respectively) 
subsamples.}
	\label{CDS}
       \end{figure}

In this subsection we analyse the distributions of SFR, SSFR, $D_n$ ($4000$), 
and $u-r$ colour for non-mergers (S1NM, S2NM, S3NM) and mergers (S1M, S2M, 
S3M) as shown in Figs.~\ref{SFS} and \ref{CDS}, whereas the median values for 
low-mass, intermediate-mass and high-mass galaxies are shown in Tables 
\ref{ML}, \ref{MI} and \ref{MH}, respectively. Since visual inspection of 
histogram figures can introduce bias, we performed a Kolmogorov-Smirnov (KS) 
statistical test (as detailed in Refs.\ \cite{hodges1958significance,harari2009kolmogorov}) 
to quantitatively assess differences between 
the distributions of mergers and non-mergers, 
taking the null hypothesis that all mergers and non-mergers group samples 
originate from the same population. A higher probability (P $\geq 0.05$) and 
low KS-statistics (close to zero) indicate that the given distributions are 
similar, while a low probability (P $<0.05$) and high KS-statistics imply 
that the given distributions are different.  Table \ref{ST} shows the 
KS statistics and P-values for the analysed subsamples. 

    \begin{table}[!h]
    	\centering
    	\caption{Median SFR, SSFR, D$_n$(4000), and $u-r$ colour for 
    		low-mass non-mergers (NM) and mergers (M). Uncertainties correspond 
    		to the 16th and 84th percentiles for each distribution.}
    	\vspace{8pt}
    	\setlength{\tabcolsep}{0.6pc}
    	\scalebox{1}{
    		\begin{tabular}{ccccccccc}
    			\toprule
    			\toprule
    			Samples  & \multicolumn{2}{c}{log$_{10}$ SFR [$M_{\odot}$/yr]} &
    			\multicolumn{2}{c}{log$_{10}$ SSFR [yr$^{-1}$]} & 
    			\multicolumn{2}{c}{D$_n$(4000)} & 
    			\multicolumn{2}{c}{$u-r$ [mag]} \\
    			\cmidrule(lr){2-3} \cmidrule(lr){4-5} \cmidrule(lr){6-7} \cmidrule(lr){8-9}
    			& NM & M & NM & M & NM & M & NM & M \\
    			{(1)} & {(2)} & {(3)} & {(4)} & {(5)} & {(6)} & {(7)} & {(8)} & {(9)} \\
    			\midrule
    			S1 & $-0.17^{+0.35}_{-0.52}$ & $0.06^{+0.41}_{-0.44}$ &
    			$-9.76^{+0.38}_{-0.43}$ & $-9.46^{+0.38}_{-0.54}$ &
    			$1.36^{+0.14}_{-0.12}$ & $1.27^{+0.13}_{-0.12}$ &
    			$1.48^{+0.48}_{-0.32}$ & $1.31^{+0.43}_{-0.27}$ \\[3pt]
    			
    			S2 & $-0.20^{+0.41}_{-0.66}$ & $0.01^{+0.30}_{-0.40}$ &
    			$-9.79^{+0.40}_{-0.59}$ & $-9.49^{+0.32}_{-0.45}$ &
    			$1.36^{+0.19}_{-0.12}$ & $1.28^{+0.12}_{-0.12}$ &
    			$1.55^{+0.67}_{-0.36}$ & $1.44^{+0.49}_{-0.31}$ \\[3pt]
    			
    			S3 & $-0.45^{+0.52}_{-1.23}$ & $-0.10^{+0.38}_{-1.37}$ &
    			$-10.05^{+0.51}_{-1.43}$ & $-9.70^{+0.46}_{-1.64}$ &
    			$1.43^{+0.45}_{-0.16}$ & $1.33^{+0.53}_{-0.14}$ &
    			$1.83^{+0.70}_{-0.50}$ & $1.59^{+0.77}_{-0.46}$ \\
    			\bottomrule
    		\end{tabular}
    	}
    	\label{ML}
    \end{table}

\begin{table}[!h]
	\centering
	\caption{Median SFR, SSFR, D$_n$(4000), and $u-r$ colour for 
		intermediate-mass non-mergers (NM) and mergers (M). Uncertainties 
		correspond to the 16th and 84th percentiles.}
	\vspace{8pt}
	\setlength{\tabcolsep}{0.6pc}
	\scalebox{1}{
		\begin{tabular}{ccccccccc}
			\toprule
			\toprule
			Samples  & \multicolumn{2}{c}{log$_{10}$ SFR [$M_{\odot}$/yr]} &
			\multicolumn{2}{c}{log$_{10}$ SSFR [yr$^{-1}$]} & 
			\multicolumn{2}{c}{D$_n$(4000)} & 
			\multicolumn{2}{c}{$u-r$ [mag]} \\
			\cmidrule(lr){2-3} \cmidrule(lr){4-5} \cmidrule(lr){6-7} \cmidrule(lr){8-9}
			& NM & M & NM & M & NM & M & NM & M \\
			{(1)} & {(2)} & {(3)} & {(4)} & {(5)} & {(6)} & {(7)} & {(8)} & {(9)} \\
			\midrule
			S1 & $-0.18^{+0.53}_{-1.33}$ & $0.11^{+0.53}_{-1.50}$ &
			$-10.50^{+0.59}_{-1.34}$ & $-10.16^{+0.49}_{-1.58}$ &
			$1.60^{+0.34}_{-0.21}$ & $1.48^{+0.36}_{-0.17}$ &
			$2.16^{+0.52}_{-0.50}$ & $2.00^{+0.71}_{-0.45}$ \\[3pt]
			
			S2 & $-0.34^{+0.72}_{-1.24}$ & $-0.08^{+0.63}_{-1.51}$ &
			$-10.67^{+0.79}_{-1.23}$ & $-10.50^{+0.78}_{-1.54}$ &
			$1.65^{+0.33}_{-0.25}$ & $1.56^{+0.39}_{-0.23}$ &
			$2.26^{+0.48}_{-0.54}$ & $2.27^{+0.49}_{-0.62}$ \\[3pt]
			
			S3 & $-1.14^{+1.23}_{-0.61}$ & $-0.97^{+1.24}_{-0.78}$ &
			$-11.44^{+1.22}_{-0.63}$ & $-11.42^{+1.31}_{-0.72}$ &
			$1.87^{+0.17}_{-0.34}$ & $1.84^{+0.18}_{-0.35}$ &
			$2.44^{+0.45}_{-0.52}$ & $2.55^{+0.33}_{-0.70}$ \\
			\bottomrule
		\end{tabular}
	}
	\label{MI}
\end{table}

\begin{table}[!h]
	\centering
	\caption{Median SFR, SSFR, D$_n$(4000), and $u-r$ colour for 
		high-mass non-mergers (NM) and mergers (M). Uncertainties 
		correspond to the 16th and 84th percentiles.}
	\vspace{8pt}
	\setlength{\tabcolsep}{0.6pc}
	\scalebox{1}{
		\begin{tabular}{ccccccccc}
			\toprule
			\toprule
			Samples  & \multicolumn{2}{c}{log$_{10}$ SFR [$M_{\odot}$/yr]} &
			\multicolumn{2}{c}{log$_{10}$ SSFR [yr$^{-1}$]} & 
			\multicolumn{2}{c}{D$_n$(4000)} & 
			\multicolumn{2}{c}{$u-r$ [mag]} \\
			\cmidrule(lr){2-3} \cmidrule(lr){4-5} \cmidrule(lr){6-7} \cmidrule(lr){8-9}
			& NM & M & NM & M & NM & M & NM & M \\
			{(1)} & {(2)} & {(3)} & {(4)} & {(5)} & {(6)} & {(7)} & {(8)} & {(9)} \\
			\midrule
			S1 & $-0.59^{+1.14}_{-0.81}$ & $-0.56^{+1.20}_{-0.87}$ &
			$-11.44^{+1.15}_{-0.85}$ & $-11.42^{+1.20}_{-0.86}$ &
			$1.87^{+0.21}_{-0.35}$ & $1.85^{+0.22}_{-0.36}$ &
			$2.54^{+0.47}_{-0.52}$ & $2.62^{+0.42}_{-0.57}$ \\[3pt]
			
			S2 & $-0.90^{+1.38}_{-0.55}$ & $-0.88^{+1.41}_{-0.56}$ &
			$-11.78^{+1.45}_{-0.58}$ & $-11.81^{+1.56}_{-0.56}$ &
			$1.94^{+0.17}_{-0.39}$ & $1.94^{+0.17}_{-0.41}$ &
			$2.63^{+0.40}_{-0.53}$ & $2.67^{+0.36}_{-0.52}$ \\[3pt]
			
			S3 & $-1.09^{+1.07}_{-0.42}$ & $-1.08^{+1.05}_{-0.46}$ &
			$-12.11^{+1.13}_{-0.39}$ & $-12.08^{+1.11}_{-0.39}$ &
			$2.03^{+0.12}_{-0.26}$ & $2.04^{+0.11}_{-0.25}$ &
			$2.72^{+0.32}_{-0.38}$ & $2.75^{+0.30}_{-0.35}$ \\
			\bottomrule
		\end{tabular}
	}
	\label{MH}
\end{table}

\begin{table}[h!]
	\centering
	\caption{The KS-statistics and their P-values as indicated in the 
brackets for SFR, SSFR, D$_n$ (4000), and $u-r$ colour.}
	\vspace{2pt}
	\setlength{\tabcolsep}{0.2pc}
	\scalebox{0.75}{
		\begin{tabular}{ccccc|cccc|cccc}
			\toprule
			\toprule
			& \multicolumn{4}{c}{Low-mass} & \multicolumn{4}{c}{Intermediate-mass}& \multicolumn{4}{c}{High-mass}\\
			\cmidrule(lr){2-5} \cmidrule(lr){6-9}\cmidrule(lr){10-13}
			Sample & SFR & SSFR & D$_n$ (4000) & $u-r$  & SFR & SSFR& D$_n$ (4000) & $u-r$& SFR & SSFR& D$_n$ (4000) & $u-r$ \\
			{(1)} & {(2)} & {(3)} & {(4)} & {(5)} & {(6)} & {(7)}& {(8)}& {(9)} & {(10)} & {(11)}& {(12)}&{(13)} \\
			\midrule
			S1 & 0.24 (5.7e-10) & 0.33 (1.5e-18) & 0.31 (1.4e-16) & 0.20 (6.2e-07)& 0.21 (1.2e-06) & 0.19 (3.1e-05) & 
			0.23 (1.3e-07)& 0.14 (6.1e-03)& 0.05 (0.683)&0.04 (0.936)& 0.04 (0.930)& 0.08 (0.210)\\[2pt]
			S2 & 0.30 (1.8e-13) & 0.28 (2.1e-11) & 0.29 (4.6e-12) & 0.13 (7.5e-03) & 0.13 (3.3e-03) & 0.10 (4.0e-02)& 
			0.21 (1.6e-02)& 0.06 (4.1e-02)& 0.02 (0.999)& 0.03 (0.818)& 0.02 (1.000)& 0.06 (0.179) \\[2pt]
			S3 & 0.28 (3.8e-04) & 0.30 (1.6e-04) & 0.31 (5.2e-05) &0.22 (1.3e-02)& 0.10 (4.8e-02) & 0.09 (6.5e-02)& 
			0.11 (4.0e-02)& 0.10 (4.7e-02)&0.03 (0.999)& 0.03 (0.987)& 0.03 (0.992)& 0.05 (0.860) \\
						\bottomrule
		\end{tabular}}
	\label{ST}
\end{table}
\subsection{Relationships Between Properties}
\begin{figure}[h!]
	\subfigure{
		\includegraphics[width=0.32\linewidth]{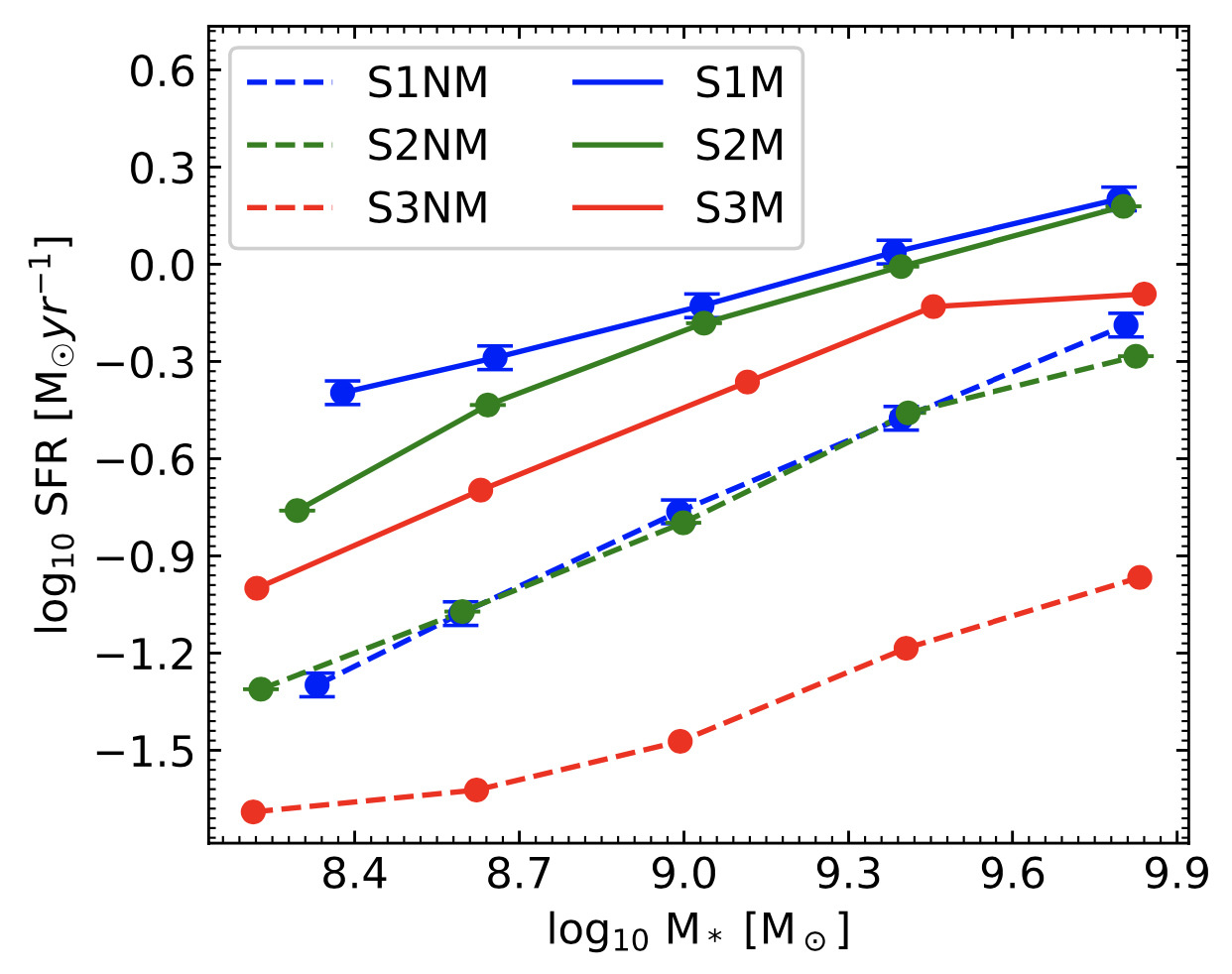}
		\label{SFLA}
	}
		\hspace{-0.2cm}
	\subfigure{
		\includegraphics[width=0.32\linewidth]{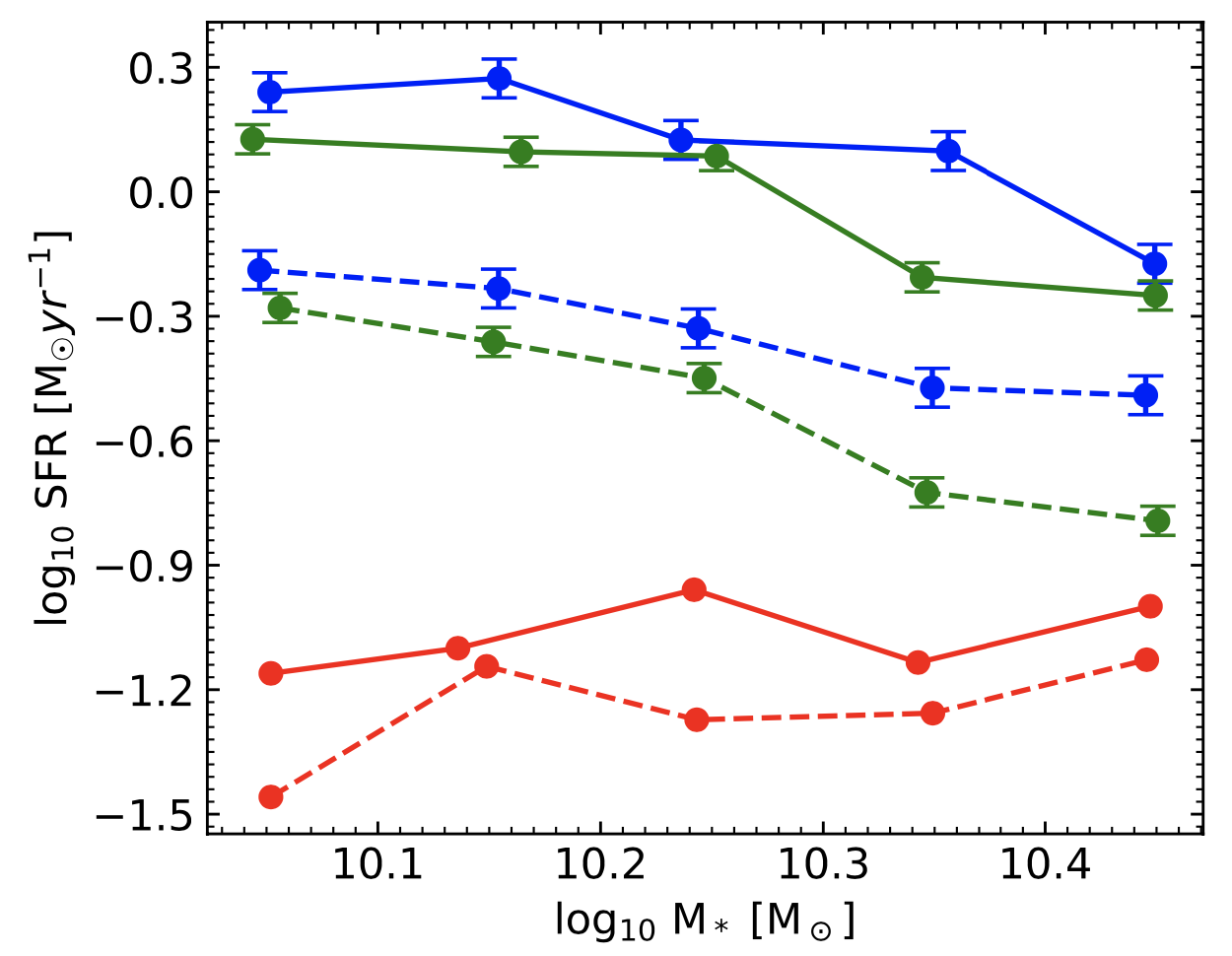}
		\label{SFIA}
	}
		\hspace{-0.2cm}
	\subfigure{
		\includegraphics[width=0.32\linewidth]{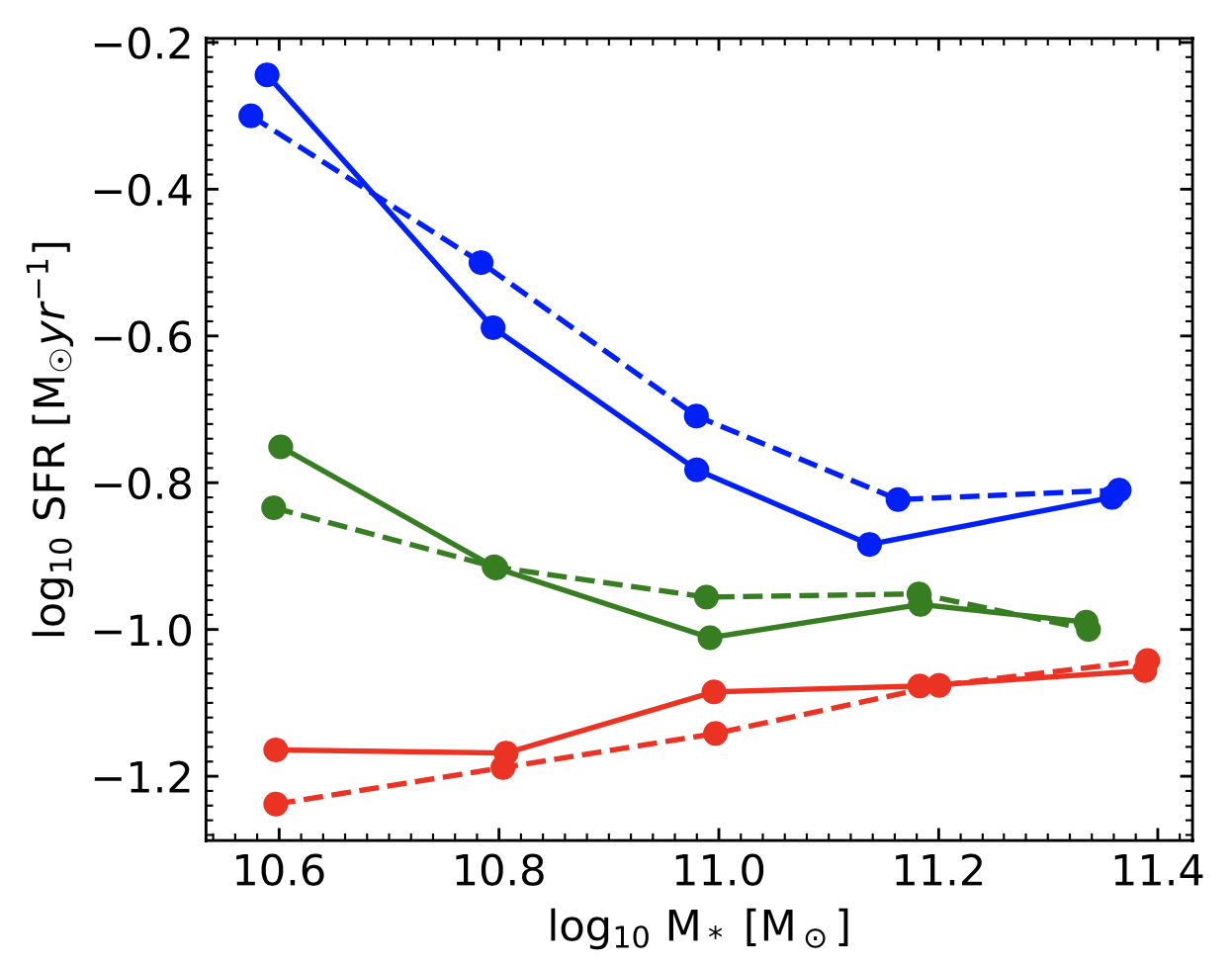}
		\label{SFHA}
	}
	\hspace{-0.2cm}
		\subfigure{
		\includegraphics[width=0.32\linewidth]{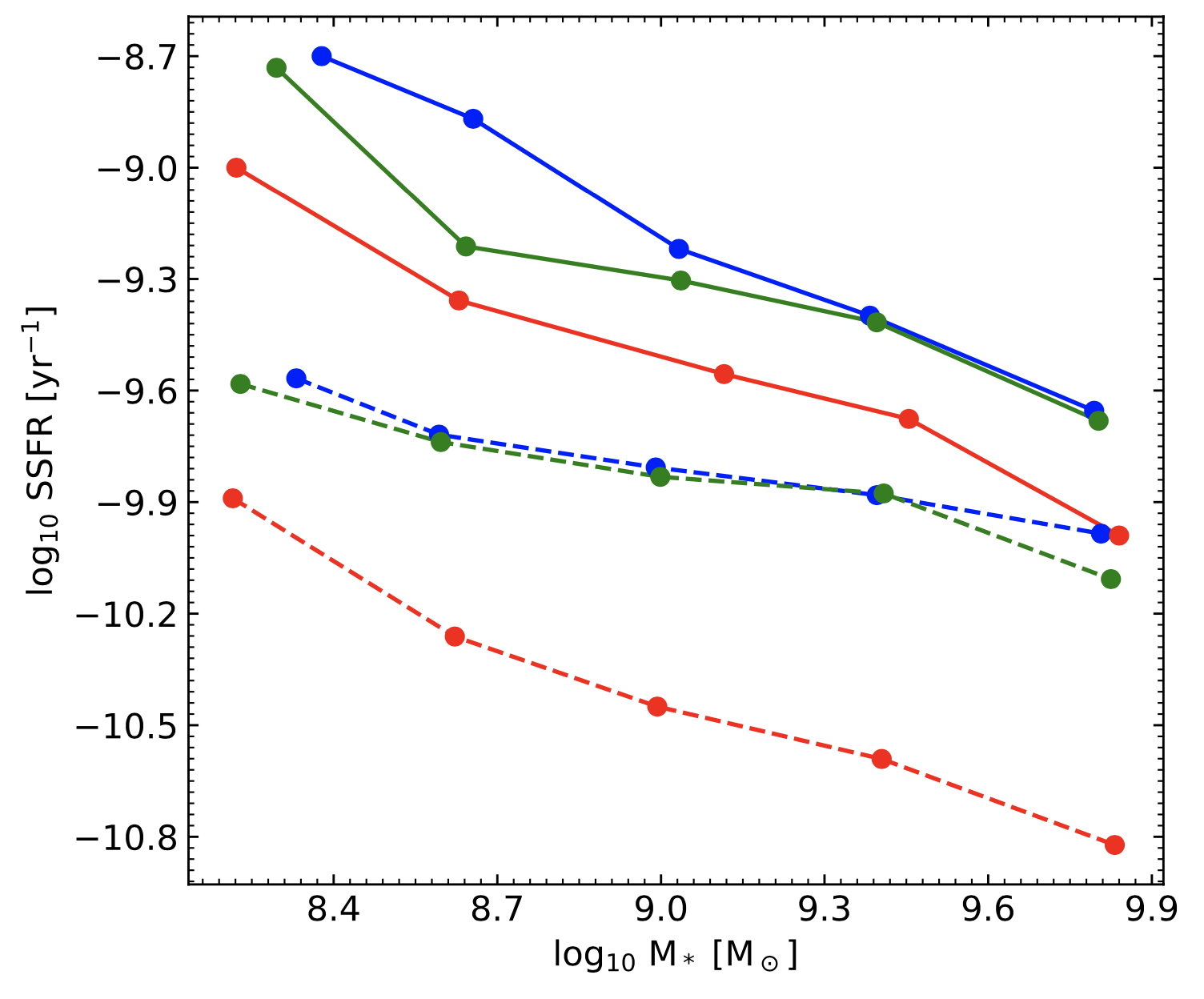}
		\label{SSFLA}
	}
		\hspace{-0.2cm}
	\subfigure{
		\includegraphics[width=0.32\linewidth]{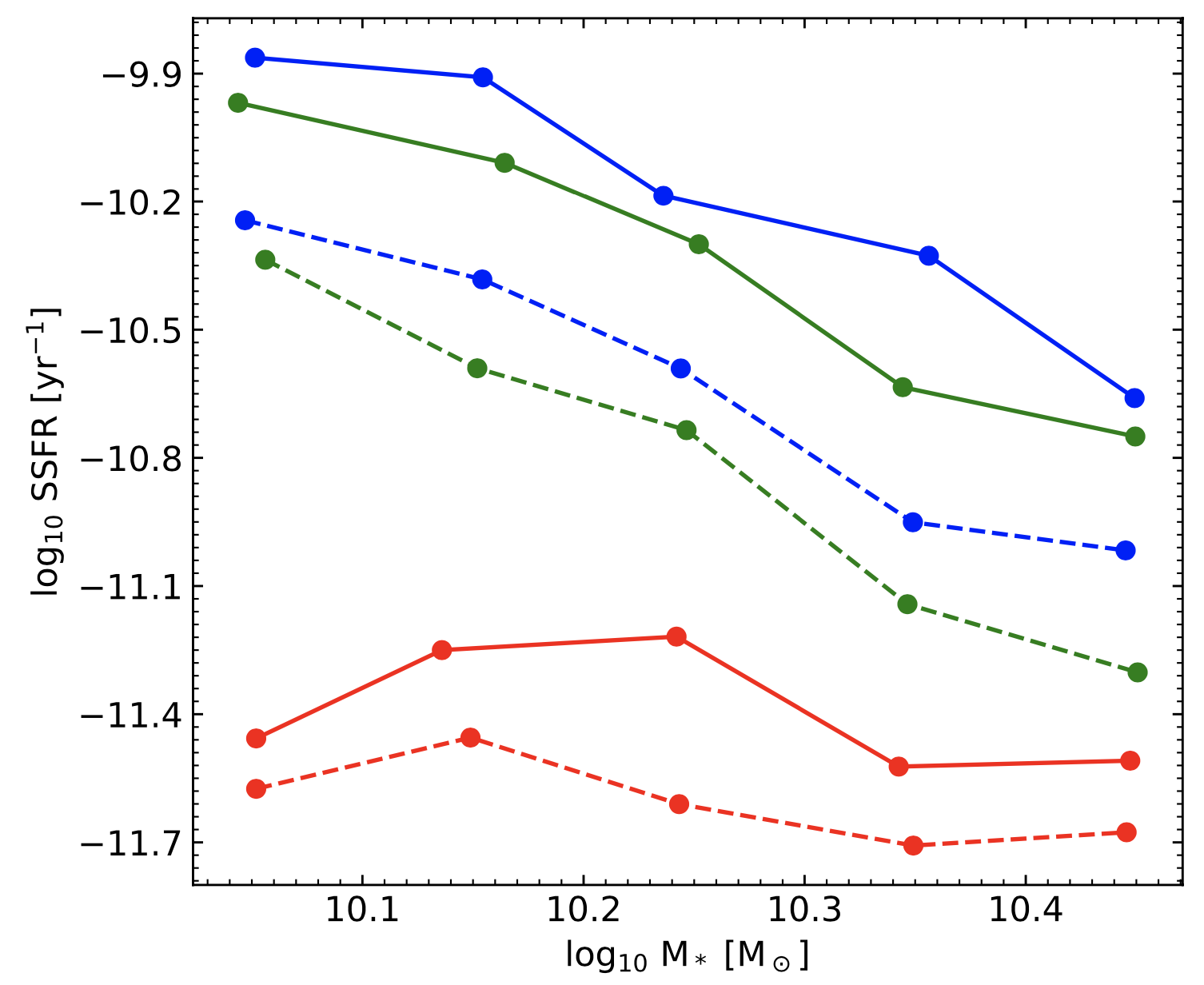}
		\label{SSFIA}
	}
		\hspace{-0.2cm}
	\subfigure{
		\includegraphics[width=0.32\linewidth]{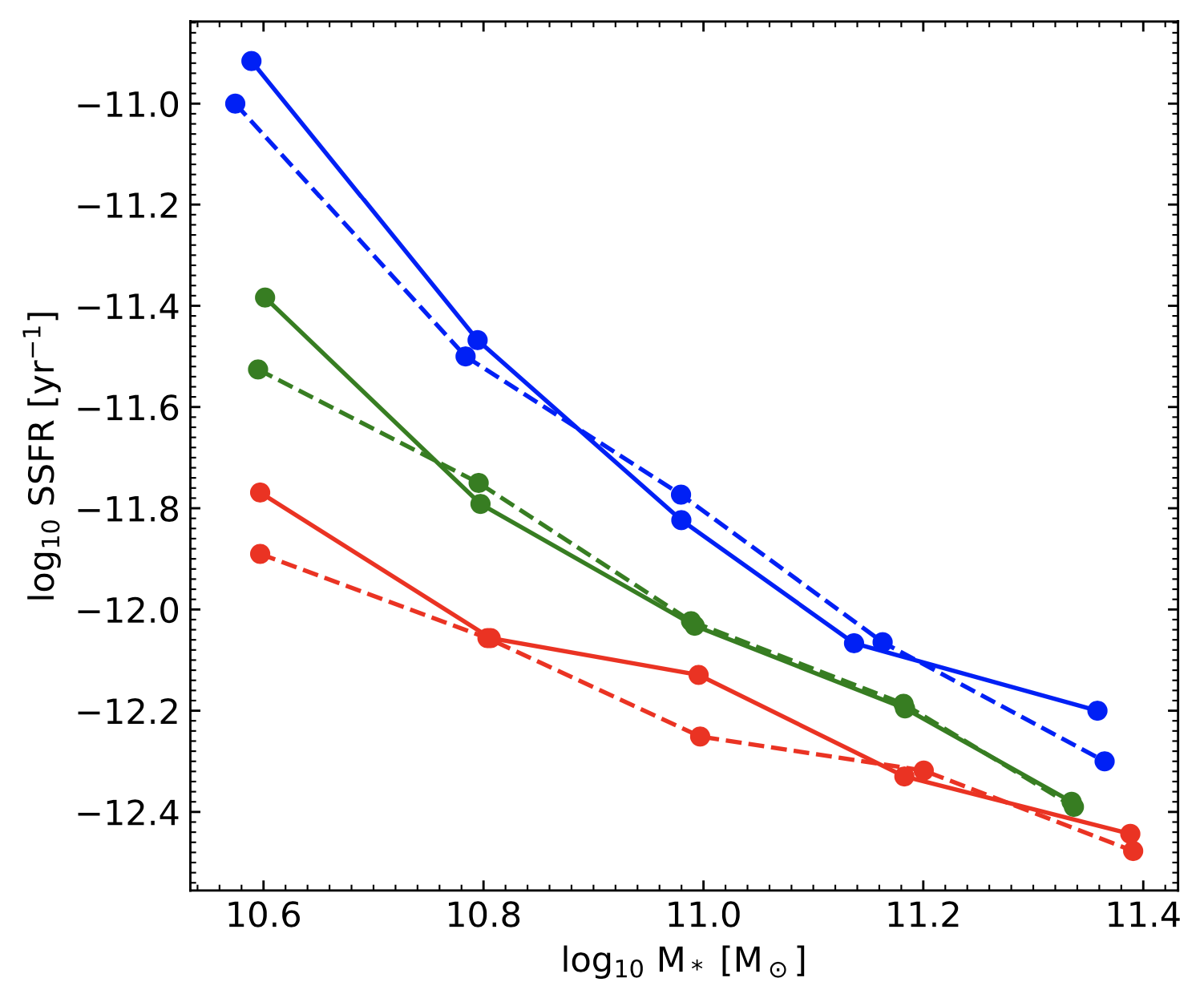}
		\label{SSFHA}
	}
	\vspace{-0.2cm}
\caption{SFR (top panels) and SSFR (bottom panels) as a function of M$\star$ 
for low-mass, intermediate-mass, and high-mass (from left to right, 
respectively) subsamples.}
	\label{SFSA}
\end{figure}

\begin{figure}[t]
	\subfigure{
		\includegraphics[width=0.31\linewidth]{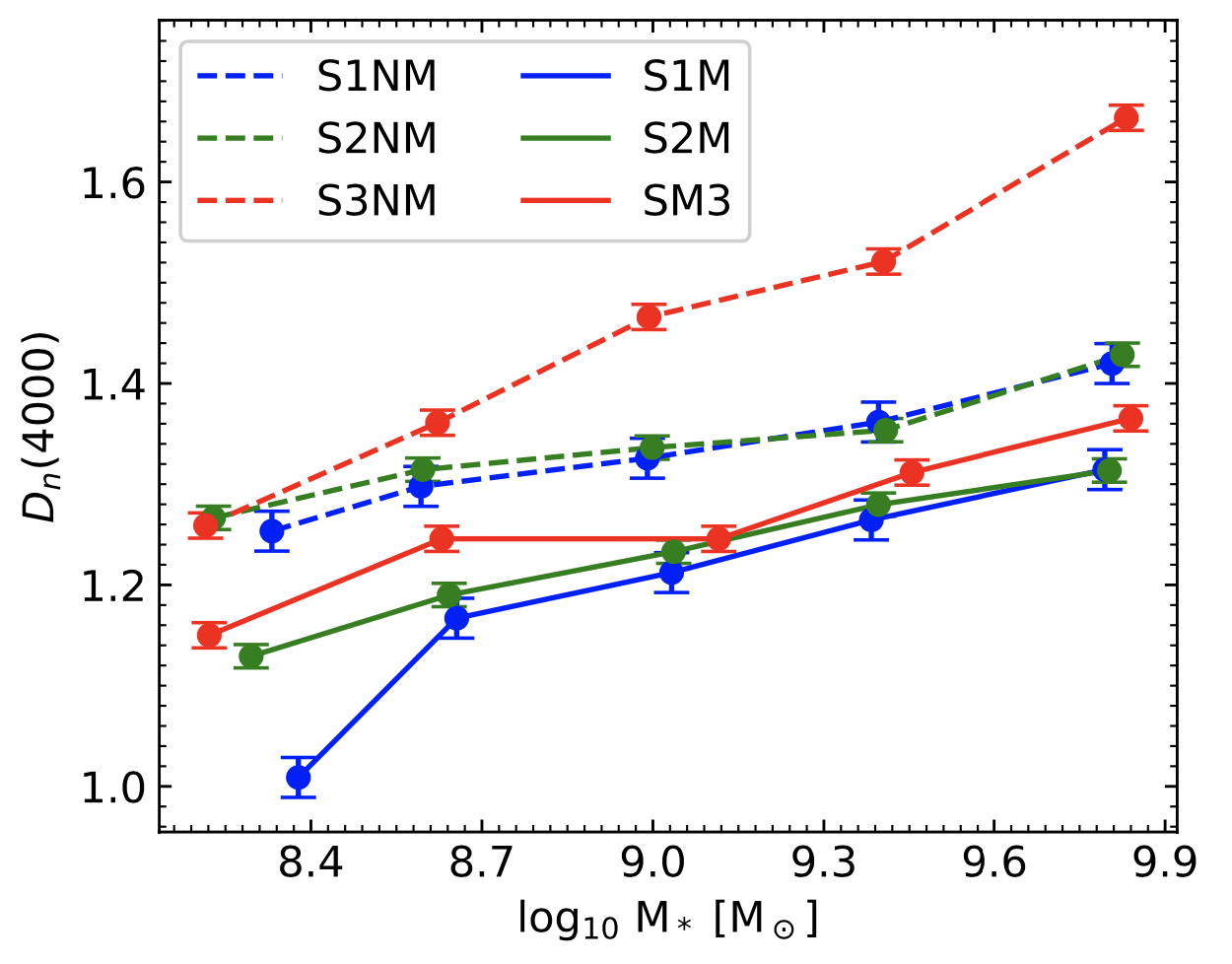}
		\label{DLA}
	}
	\hspace{-0.2cm}
	\subfigure{
		\includegraphics[width=0.31\linewidth]{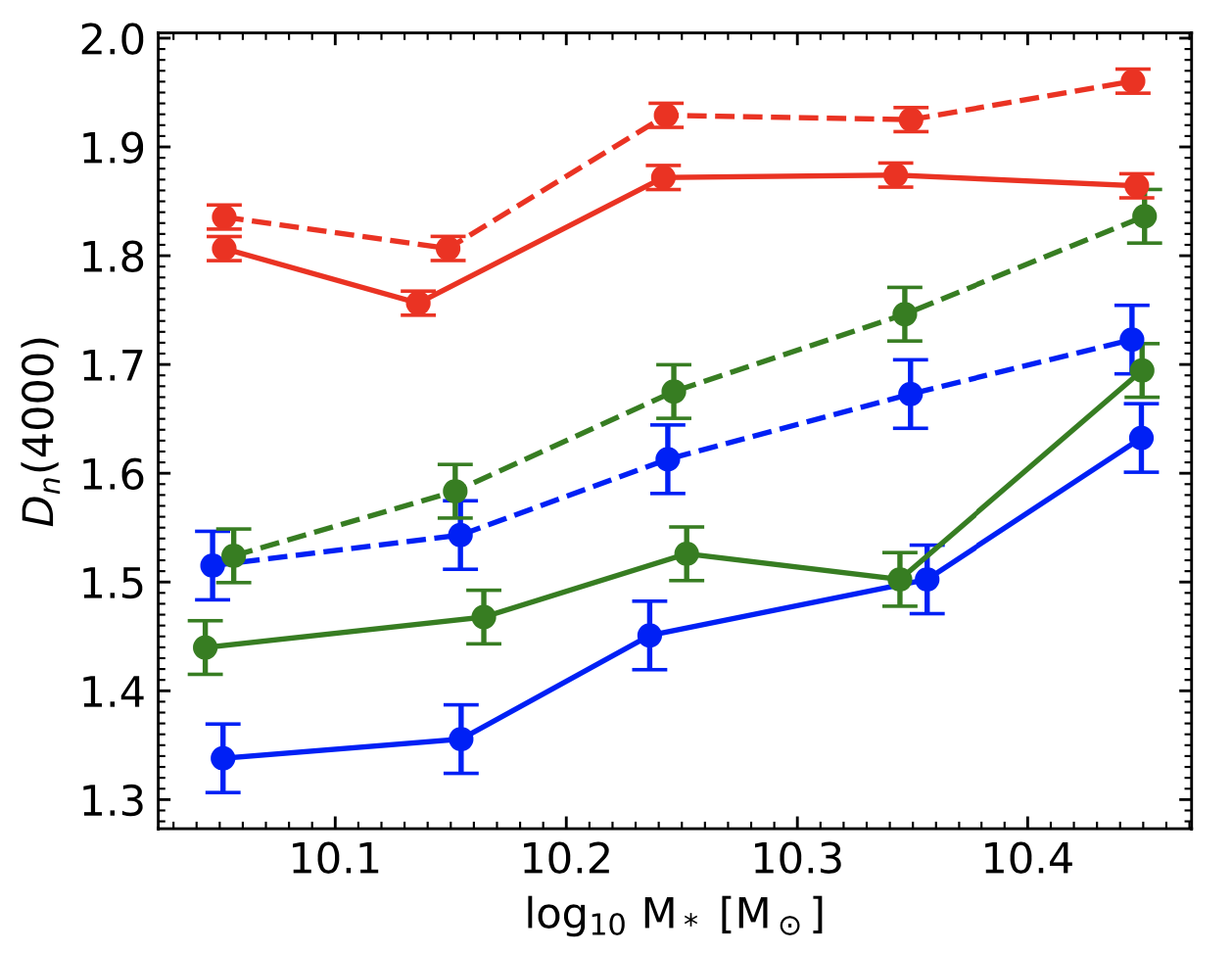}
		\label{DIA}
	}
	\hspace{-0.2cm}
	\subfigure{
		\includegraphics[width=0.31\linewidth]{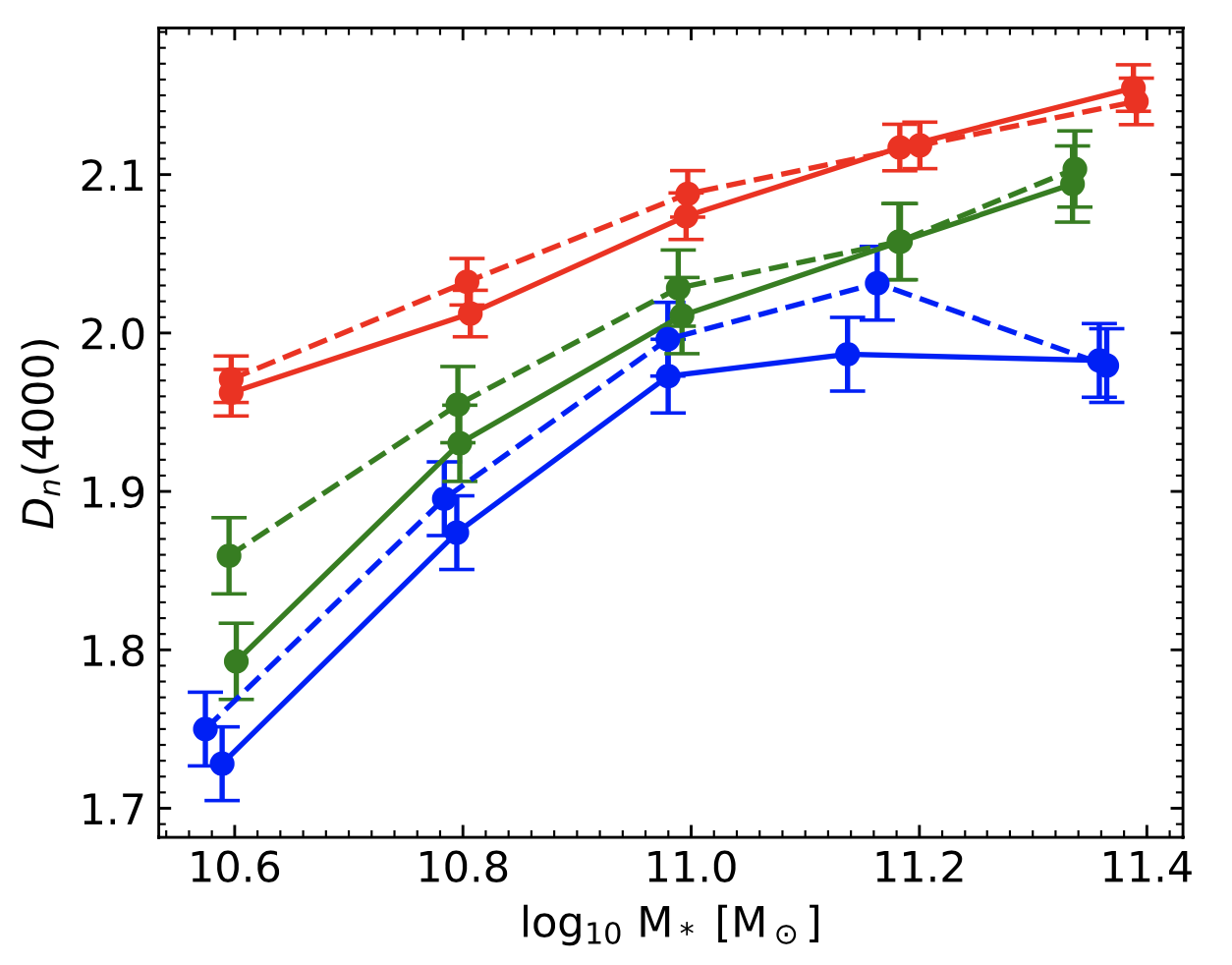}
		\label{DHA}
	}
	\hspace{-0.2cm}
		\subfigure{
		\includegraphics[width=0.31\linewidth]{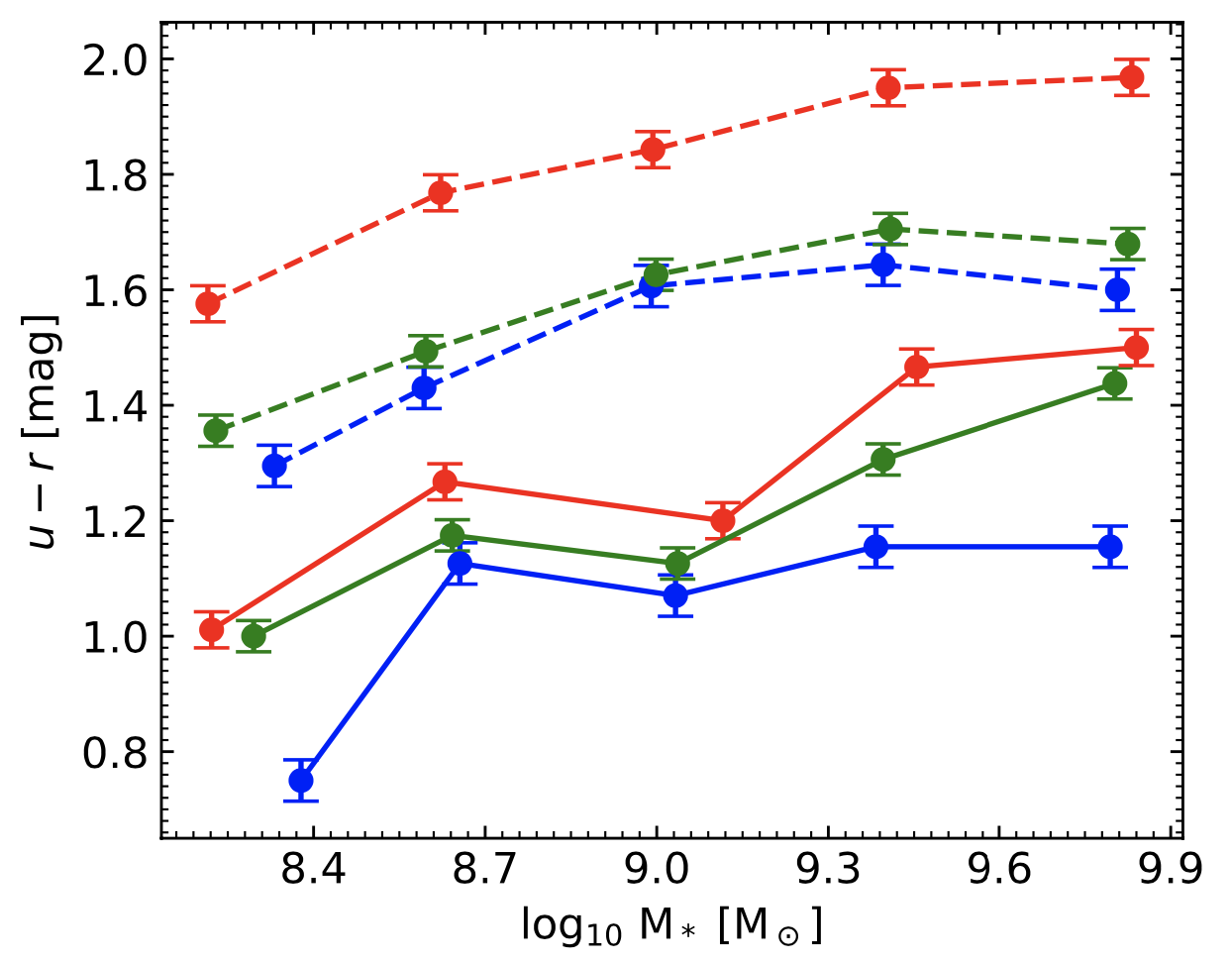}
		\label{CLA}
	}
	\hspace{-0.2cm}
	\subfigure{
		\includegraphics[width=0.31\linewidth]{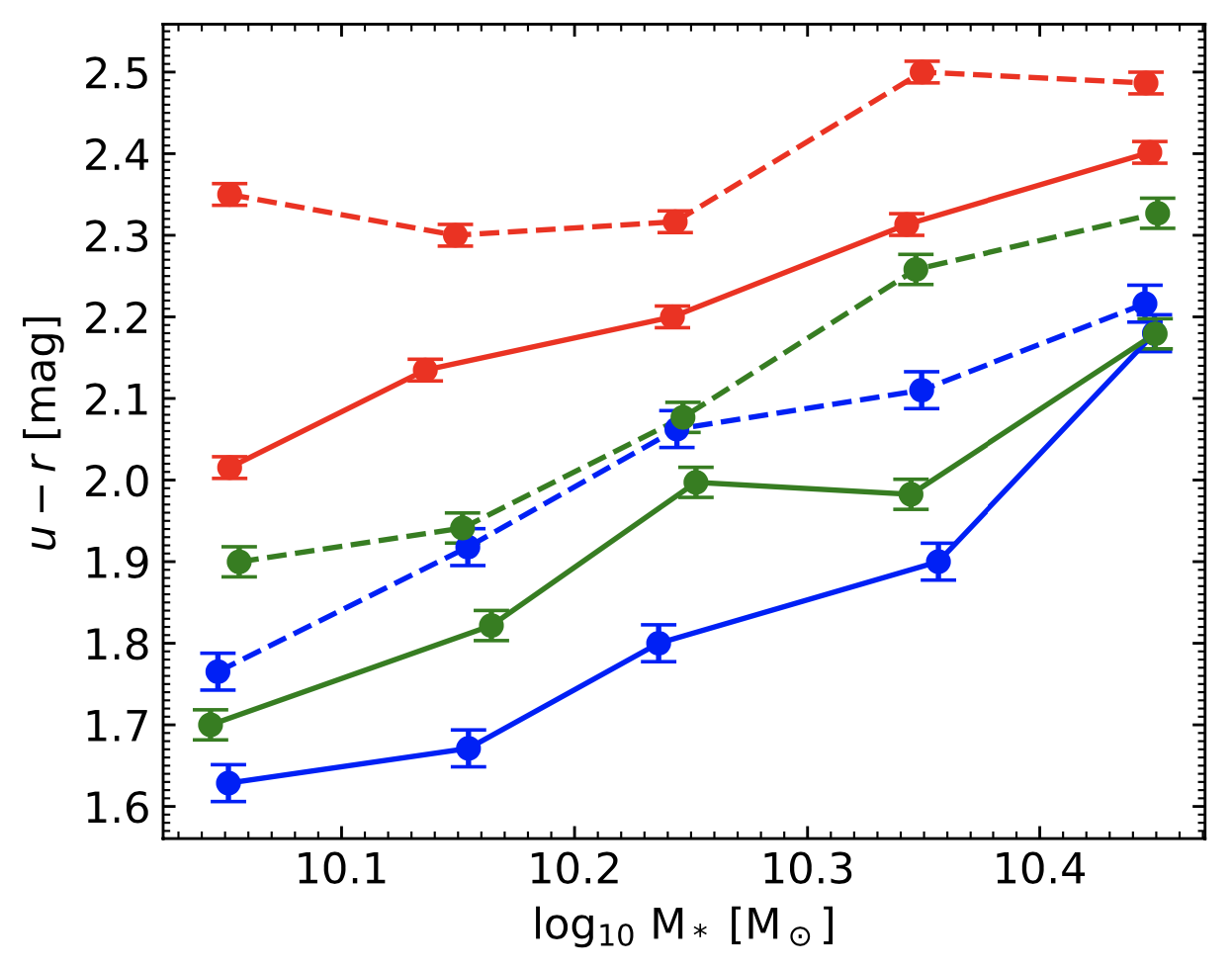}
		\label{CIA}
	}
	\hspace{-0.2cm}
	\subfigure{
		\includegraphics[width=0.31\linewidth]{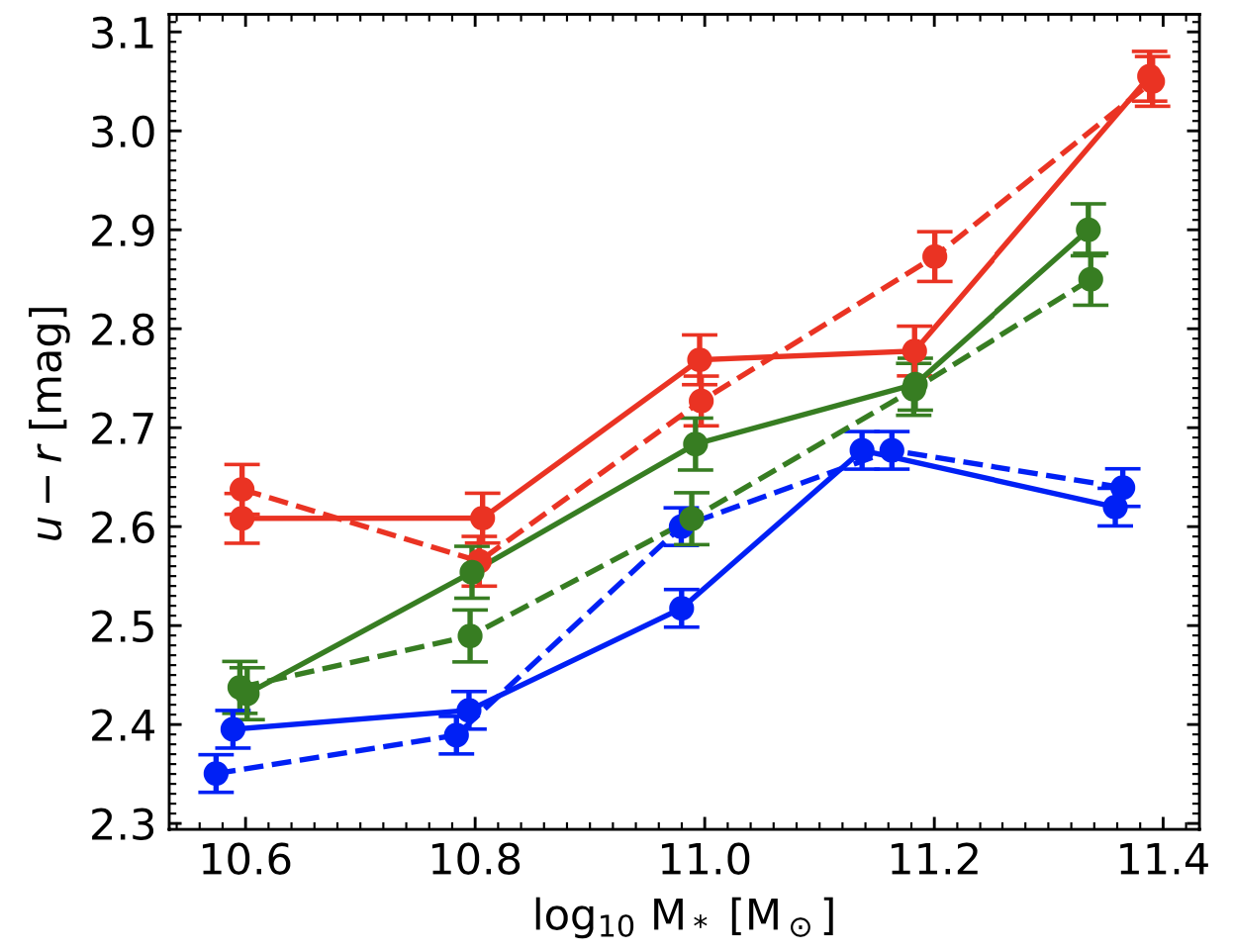}
		\label{CHA}
	}
	\vspace{-0.2cm}
\caption{D$_n$ (4000) (top panels) and $u-r$ colour (bottom panels) as a 
function of M$\star$ for low-mass, intermediate-mass, and high-mass (from left 
to right hand panel, respectively) subsamples.} 
	\label{CDSA}
\end{figure}

In this subsection, we relate the variations of SFR, SSFR, D$_n$ (4000) and 
$u-r$ colour with stellar mass for each subsample of galaxies by further 
dividing them into five categories for the mass binning sizes of 
$0.4$, $0.1$, and $0.2$ for low-mass, intermediate-mass and high-mass 
subsamples respectively as shown in Figs.~\ref{SFSA} and \ref{CDSA}.  
\section{Discussion}\label{secV}
From Fig.~\ref{SFS} and columns (2), (3), (4) and (5) of Tables \ref{ML}, 
\ref{MI} and \ref{MH}, it is observed that the SFR and SSFR decrease from 
low-mass to high-mass galaxies for both mergers and non-mergers. Also, the SFR
and SSFR decrease with the increase of group richness as the decrease is 
observed from S1 to S3 for low-mass, intermediate-mass and high-mass 
subsamples, however the difference is maximum for low-mass and minimum for 
high-mass galaxies. Mergers possess higher SFR and SSFR when compared to 
non-mergers, supporting the recent studies by Refs.\ 
\cite{renaud2022merger,shah2022investigating,hani2020interacting,
pearson2019effect,knapen2015interacting,thorp2024all}. Nonetheless, in this 
study, we observe that the difference is significant for low-mass galaxies, 
while for high-mass galaxies the difference is negligible, supporting the 
results from Ref.\ \cite{li2023subtle}. As we observe the values in Tables 
\ref{ML}, \ref{MI} and \ref{MH}, we found from the statistics test in Table 
\ref{ST} the average KS statistics and P-values used to compare the 
distribution of SFR between mergers and non-mergers for low-mass sample as 
$0.27$, $\num{1.27e-04}$; for intermediate-mass as $0.15$, $\num{0.16}$ and for 
high-mass as $0.03$, $\num{0.89}$. For the case of SSFR, the average KS 
statistics, P-values are $0.30$, $\num{5.3e-05}$ for low-mass; $0.13$, 
$\num{0.23}$ for intermediate-mass; $0.03$, $\num{0.91}$ for high-mass 
galaxies. Based on these statistics, we are confident to state that the 
difference in the distribution of SFR and SSFR between mergers and 
non-mergers depends on stellar mass, with a significant difference for low-mass 
and a negligible difference for high-mass. From the top left and bottom 
left panels of Fig.~\ref{SFSA}, it is observed that mergers have high SFR and 
SSFR when compared to non-mergers for low-mass galaxies, supporting the 
observations from Tables \ref{ML}, \ref{MI} and \ref{MH}. Another insight from 
these figures is that the SFR decreases with the increase in group richness, as 
it is observed that S3 has a lower SFR, SSFR than S2 and S1 when the merging 
status is kept constant. For the intermediate-mass, it is observed from middle 
top to bottom panels of Fig.~\ref{SFSA} that the rich group galaxies have 
minimized SFR and SSFR when compared to poor group galaxies when the merging 
status is kept constant. Again from the top and bottom right panels of 
Fig.~\ref{SFSA} there is no difference in SFR and SSFR of mergers and 
non-mergers, although again, as the group richness increases the SFR and SSFR 
decrease.

From Fig.~\ref{CDS} and columns (6), (7), (8) and (9) of Tables \ref{ML}, 
\ref{MI} and \ref{MH}, we observed that the D$_n$ (4000) and $u-r$ colour 
increases from low-mass to high-mass galaxies for both mergers and 
non-mergers. Similarly, the increase of these properties with group richness 
is observed, where S1 has lower D$_n$ (4000) and $u-r$ colour when compared 
to S3. Mergers possess lower D$_n$ (4000), indicating a younger stellar 
population and  bluer in colour when compared to non-mergers for low-mass 
galaxies, supporting the findings from Ref.~\cite{sureshkumar2024galaxy}, 
however, for high-mass the difference between mergers and non-mergers in 
both D$_n$ (4000) and $u-r$ is very small. From Table \ref{ST} it is observed 
that the distribution D$_n$ (4000) for low-mass galaxies is stronger having 
the average KS test, P-values of $0.3$, $\num{1.7e-05}$ when compared to 
$0.18$, $\num{5.5e-03}$ and $0.03$, $0.974$ of intermediate-mass and 
high-mass galaxies, respectively. Since the KS statistics is closer to zero 
and the P-value is greater than $\num{5e-2}$ for high-mass, while the opposite 
case is observed in low-mass galaxies, this indicates that mergers' 
distribution is different from non-mergers for low-mass, while for high-mass 
galaxies the distributions of D$_n$ (4000) for mergers and non-mergers are 
similar. Moreover, the $u-r$ colour distributions indicate that mergers and 
non-mergers have similar distributions for high-mass galaxies with average 
KS statistics and P-value of $0.19$ and $\num{0.42}$, while for low-mass 
galaxies the distributions are different with KS statistics and p-value of 
$0.18$ and $\num{0.0032}$, respectively. For intermediate-mass, the KS 
statistics and p-value of $0.1$ and $\num{0.29}$ in average, respectively, 
were obtained, indicating similar distributions. Fig.~\ref{CDSA} shows the 
D$_n$ (4000) as a function of M$\star$. It should be kept in mind that 
higher values of D$_n$ (4000) indicate an older stellar population while lower 
values indicate a younger stellar population, thus from Fig.~\ref{CDSA} it is
clear that the non-mergers form the older stellar population than the mergers 
for the low-mass sample. We observe a clear trend between the group richness 
and the D$_n$ (4000), where the  D$_n$ (4000) increases with the increase 
of group richness when the merging status is kept constant. No matter the 
stellar mass, it proves that the galaxies arrange themselves in companions as 
they becomes older. Further, we see the absence of a significant difference 
between mergers and non-mergers D$_n$ (4000) for high-mass galaxies. In 
Fig.~\ref{CDSA}, the $u-r$ colour as a function of M$\star$ for low-mass, 
intermediate-mass and high-mass subsamples is also shown, and it is observed 
that rich group galaxies show a clear higher $u - r$ colour, indicating redder,
while poor group galaxies have smaller $u - r$ colour indicating bluer no 
matter the stellar mass. For the low-mass sample, mergers are bluer than 
non-mergers. Furthermore, for high-mass samples, rich group galaxies are 
redder than poor group galaxies. We again observe the absence of significant 
difference in $u-r$ colour between mergers and non-mergers for high-mass 
galaxies indicating, that galaxy colour is influenced by merging status, 
group richness and the stellar mass.

The results presented above demonstrate clear trends with stellar mass, 
merging status, and group richness. While the preceding analysis has been 
largely descriptive, these behaviours can be physically motivated within 
current frameworks of galaxy evolution. The weak difference between mergers 
and non-mergers in high-mass galaxies may be explained by the fact that 
such systems often reside in massive, virialized halos where gas reservoirs 
are already depleted or shock-heated to high temperatures. In these 
environments, mergers are less effective in triggering star formation, as 
the supply of cold gas is limited. This is consistent with the picture of 
quenching by hot halos and AGN feedback, where the impact of mergers is 
reduced compared to low-mass galaxies that remain gas-rich.

The decline in SFR and SSFR with group richness can be interpreted as an 
environmental effect, where processes such as ram-pressure stripping, 
strangulation, and tidal interactions remove or heat the cold gas supply. 
The stronger impact on low-mass galaxies is consistent with their shallower 
gravitational potentials, making them more susceptible to gas loss in dense 
environments. Meanwhile, the older stellar populations and redder colours 
observed in rich groups indicate accelerated quenching histories, likely 
linked to both pre-processing in subgroups and environmental quenching in 
the main halo. Overall, these results support a scenario in which mergers are an important 
channel of star formation enhancement primarily for gas-rich, low-mass 
systems, while in massive galaxies and rich environments, environmental 
quenching and halo processes dominate, diminishing the role of mergers.

\section{Summary and conclusion}\label{secVI}
This study involves the use of a luminous volume-limited sample constructed 
from the value added catalogue of SDSS DR 12 by Ref.\ \cite{tempel2017merging} 
to investigate the influence of group richness on the distributions and 
relationships of mergers' and non-mergers' properties for the redshift range 
covering $z\leq0.1$. The galaxies were classified in three mass-limited 
subsamples, restricting $z\geq0.03$ for completeness of low-mass 
($8\leq \log_{10} $M$\star< 10$), intermediate-mass 
($10\leq \log_{10} $M$\star< 10.5$) and high-mass 
($10.5\leq \log_{10} $M$\star< 11.5$) samples resulting to $3051$, $2439$ and 
$2833$ galaxies, respectively as shown in Fig.~\ref{MZ}. Mergers were obtained 
from the GZP catalogue as detailed in Ref.\ \cite{darg2010galaxy}, with a 
total number of $470$ (low-mass), $526$ (intermediate-mass) and $1118$ 
(high-mass) mergers were obtained. Similarly, for non-mergers we obtained a 
total number of $2581$ (low-mass), $1913$ (intermediate-mass) and $1715$ 
(high-mass) galaxies. To ensure that the analyses are not stellar mass or 
redshift influenced, their distributions between mergers and non-mergers for 
all subsamples were analysed as shown in Fig.~\ref{MZD}. Aiming at comparing 
mergers and non-mergers having different group richness, the information on 
galaxies' group membership detailed in Ref.\ \cite{tempel2017merging} was 
employed as discussed in Section \ref{qn}. The mergers (S1M, S2M, S3M) and 
non-merges (S1NM, S2NM, S3NM) subsamples were constructed based on group 
richness ($N_{rich}$) that is S1 ($N_{rich} = 1$), S2 ($2\leq N_{rich} 
\leq5$) and S3 ($6 \leq N_{rich} \leq50$) are arranged from low to high 
group richness, respectively to obtain the number of galaxies as detailed 
in Section \ref{qn}. The comparison of distributions for mergers and 
non-mergers SFR, SSFR, D$_n$ (4000) and $u-r$ colour is as shown in 
Figs.~\ref{SFS} and \ref{CDS}. Tables \ref{ML}, \ref{MI} and \ref{MH} show 
the medians with their corresponding percentiles for low-mass, 
intermediate-mass and high-mass galaxies, respectively. To observe if there 
is any existence of difference between mergers and non-mergers distributions, 
the KS statistics were employed as shown in Table \ref{ST}. Further, the 
variations of SFR, SSFR, $D_n$ (4000) and $u-r$ colour with stellar mass were 
analysed in which the galaxies in each subsample were further classified into 
mass binning sizes of $0.4$, $0.1$, and $0.2$ for low-mass, intermediate-mass 
and high-mass, respectively as shown in Figs.~\ref{SFSA} and \ref{CDSA}. With
all these analyses, in addition to previously established findings, this 
study exposed the following insight:
\begin{itemize}
	\item [1.] There is a significant difference in the distribution of 
        M$\star$, SFR, SSFR, D$_n$ (4000) and $u-r$ colour between mergers 
        and non-mergers of low-mass galaxies, while for high-mass galaxies 
        there is a weak difference.
	\item [2.] A clear trend was observed between the group richness  
        ($N_{rich}$) and SFR, SSFR, wherein the SFR, SSFR 
	decrease with group richness when the merging status is kept constant,
        however, the relations are independent of mass.
	\item [3.] Mergers have higher SFR and SSFR than non-mergers for 
	low-mass galaxies when the group richness is kept constant, while for 
	high-mass rich group galaxies have lower SFR and SSFR when the merging 
        status is kept constant. 
	\item [4.] The galaxies in rich groups resemble the older stellar 
        population and are redder exhibited by higher D$_n$ (4000) and $u-r$ 
        colour when compared to poor groups, when the merging status is kept 
        constant. Again the relation is independent of mass.
	\item [5.] Mergers resemble young stellar populations, and are bluer 
        than non-mergers for low-stellar mass galaxies when the group richness 
        is kept constant, while for high-mass, poor group galaxies resemble 
        young stellar populations when merging status is kept constant. 
	\end{itemize}
This study revealed that the properties' distributions and relationships 
between mergers and non-mergers in the local Universe are influenced by stellar 
mass and group richness. In the future, the study will be extended to higher 
redshifts, aiming to investigate further the redshift influence on the observed 
trends.

\section*{Acknowledgments} PP acknowledges support from The Government of 
Tanzania through the India Embassy, Mbeya University of Science and Technology 
(MUST) for Funding for doing Ph.D.~in India. UDG is thankful to the 
Inter-University Centre for Astronomy and Astrophysics (IUCAA), Pune, India 
for the Visiting Associateship of the institute. Funding for SDSS-III has been 
provided by the Alfred P.~Sloan Foundation, the Participating Institutions, 
the National Science Foundation, and the U.S.~Department of Energy Office of 
Science. The SDSS-III website is http://www.sdss3.org/. SDSS-III is managed by 
the Astrophysical Research Consortium for the Participating Institutions of 
the SDSS-III Collaboration including the University of Arizona, the Brazilian 
Participation Group, Brookhaven National Laboratory, Carnegie Mellon 
University, the University of Florida, the French Participation Group, the 
German Participation Group, Harvard University, the Instituto de Astrofisica 
de Canarias, the Michigan State/Notre Dame/JINA Participation Group, Johns 
Hopkins University, Lawrence Berkeley National Laboratory, Max Planck 
Institute for Astrophysics, Max Planck Institute for Extraterrestrial Physics, 
New Mexico State University, New York University, Ohio State University, 
Pennsylvania State University, University of Portsmouth, Princeton University, 
the Spanish Participation Group, University of Tokyo, the University of Utah, 
Vanderbilt University, the University of Virginia, the University of 
Washington, and Yale University.

\end{document}